\RequirePackage{fix-cm}
\documentclass[smallextended]{svjour3}       
\smartqed  
\usepackage{graphicx}
\usepackage{enumitem}

\usepackage{commath}
\usepackage{ragged2e}
\usepackage{multirow}
\usepackage{amsmath}
\usepackage{mathtools}
\usepackage{url}
\usepackage{balance}
\usepackage{booktabs}
\usepackage{listings}
\usepackage{algorithm}
\usepackage{algpseudocode}
\usepackage{caption}
\usepackage{xspace}
\usepackage{url}
\usepackage{hyperref}
\usepackage[switch]{lineno}
\usepackage[authoryear]{natbib}
\usepackage{adjustbox}

\usepackage{varwidth}

\usepackage{amsmath}
\usepackage{tikz}
\usetikzlibrary{shapes.geometric, arrows.meta, positioning, calc}
\usepackage{booktabs}
\usepackage{caption}

\usepackage{tabularx} 
\usepackage{pifont}

\tikzstyle{stage} = [rectangle, minimum width=2.6cm, minimum height=0.9cm, text centered, draw=black, fill=blue!10]
\tikzstyle{optional} = [stage, fill=gray!15, dashed]
\tikzstyle{arrow} = [thick, ->, >=Stealth]

\usepackage[most]{tcolorbox}
\usepackage{soul}

\usepackage{tikz}
\usetikzlibrary{arrows.meta, positioning}

\usepackage{tikz-qtree} 
\usepackage{pgf-pie}
\usetikzlibrary{positioning}
\usepackage{threeparttable}  

\usepackage[table]{xcolor}

\usepackage{booktabs}  
\usepackage{array}     
\usepackage{lscape}
\usepackage{rotating}
\usepackage{graphicx}    
\usepackage{subcaption}
\usepackage[font=small,labelfont=bf]{caption}

\usepackage[htt]{hyphenat}
\usepackage{ulem}
\setul{0.5ex}{0.2ex}  

\usepackage{float} 
\usetikzlibrary{shapes.geometric, arrows.meta, positioning, decorations.pathmorphing}
\usetikzlibrary{decorations.pathreplacing}

\tikzstyle{stage} = [rectangle, draw=black, fill=gray!10, text width=4.5cm, text centered, minimum height=1.2cm]
\tikzstyle{arrow} = [thick,->,>=stealth]

\newif\ifDEBUG
\DEBUGfalse
\DEBUGtrue

\ifDEBUG
  \newcommand{\yuan}[1]{\textcolor{red}{[Yuan: #1]}}
  \newcommand{\jerin}[1]{\textcolor{magenta}{{\itshape [Jerin: #1]}}}
  \newcommand{\JD}[1]{\textcolor{orange}{{\itshape [Jamie: #1]}}}
  \newcommand{\WJ}[1]{\textcolor{purple}{{\itshape [Wenxin: #1]}}}
\else
  \newcommand{\yuan}[1]{}
  \newcommand{\jerin}[1]{}
  \newcommand{\JD}[1]{}
  \newcommand{\WJ}[1]{}
\fi

\newcommand{\code}[1]{{\small\texttt{#1}}\xspace}

\newcommand{\new}[1]{\textcolor{black}{#1}}

\newcommand{\rqone}{\textit{How do open-source projects structure and document their PTM dependencies?}}
\newcommand{\rqtwo}{\textit{What are the stages and their organization in the \new{reuse} pipeline of PTM-based open source projects?}}

\newcommand{\rqthree}{\textit{How do PTMs interact with other learned components?}}

\newcommand{\ie}{\textit{i.e.,}\xspace}
\newcommand{\eg}{\textit{e.g.,}\xspace}
\newcommand{\etal}{\textit{et al.}\xspace}

\newcommand{\PreliminaryTotalProjects}{{650}\xspace}

\newcommand{\TotalProjects}{{401}\xspace}

\newcommand{\ProjectsMultiPTM}{{52.6\%}\xspace}

\newcommand{\ProjectsCentralizedDocumentation}{{21.2\%}\xspace}

\newtcbtheorem{Summary}{\bfseries Summary of RQ}{enhanced,
	coltitle=black,
	top=0.1in,
	attach boxed title to top left=
	{xshift=1.5em,yshift=-\tcboxedtitleheight/2},
	boxed title style={size=small,colback=lightgray}
}{summary}

\newtcbtheorem{Summary_data_charac}{Summary}{enhanced,
	coltitle=black,
	top=0.1in,
	attach boxed title to top left=
	{xshift=1.5em,yshift=-\tcboxedtitleheight/2},
	boxed title style={size=small,colback=gray}
}{Summary_data_charac}

\newtcolorbox[auto counter]{summary}[1][]{title={\bfseries Summary},enhanced,
	coltitle=black,
	top=0.17in,
	attach boxed title to top left=
	{xshift=1.5em,yshift=-\tcboxedtitleheight/2},
	boxed title style={size=small,colback=lightgray},#1}

\newtcolorbox[auto counter]{summary_PQ}[1][]{title={\bfseries Summary of Preliminary Study},enhanced,
	coltitle=black,
	top=0.17in,
	attach boxed title to top left=
	{xshift=1.5em,yshift=-\tcboxedtitleheight/2},
	boxed title style={size=small,colback=lightgray},#1}

\newtcolorbox{TColorFindingsBox}[1]{%
  enhanced,
  coltitle=black,
  width=0.99\linewidth,
  top=0.17in,
  attach boxed title to top left={xshift=1.5em, yshift=-\tcboxedtitleheight/2},
  varwidth boxed title=0.83\linewidth, 
  boxed title style={size=small, colback=lightgray},
  fonttitle=\bfseries,
  title={#1}
}

\newcommand{\FindingsBox}[2]{%
  \begin{TColorFindingsBox}{#1}#2\end{TColorFindingsBox}
}

\begin{document}

\title{Software Dependencies 2.0: An Empirical Study of Reuse and Integration of Pre-Trained Models in Open-Source Projects}

\newcommand{\authorNote}{\textsuperscript{*}}

\authorrunning{Yasmin, Jiang, Davis, and Tian}
\titlerunning{Software Dependencies 2.0}

\author{
  Jerin Yasmin \and
  Wenxin Jiang\authorNote \and
  James C. Davis \and
  Yuan Tian
}

\begingroup
  \renewcommand\thefootnote{}
  \footnotetext{*Work primarily completed during his Ph.D. at Purdue University.}
  \addtocounter{footnote}{-1}
\endgroup



\institute{
  Jerin Yasmin \at
    Queen's University, Kingston, ON, Canada \\
    \email{jerin.yasmin@queensu.ca}
  \and
  Wenxin Jiang \at
    Socket Inc., Wilmington, DE, USA \\
    \email{wenxin@socket.dev}
  \and
    James C. Davis \at
    Purdue University, West Lafayette, IN, USA \\
    \email{davisjam@purdue.edu}
  \and
    Yuan Tian \at
    Queen's University, Kingston, ON, Canada \\
    \email{y.tian@queensu.ca} 
}

\date{Received: date / Accepted: date}

\maketitle

\begin{abstract}
\noindent\emph{Context:} Pre-trained models (PTMs) are machine learning models that have been trained in advance, often on large-scale data, and can be reused for new tasks, thereby reducing the need for costly training from scratch.  Their widespread adoption introduces a new class of software dependency, which we term \textit{Software Dependencies 2.0}, extending beyond conventional libraries to learned behaviors embodied in trained models and their associated artifacts. The integration of PTMs as software dependencies in real projects remains unclear, potentially threatening maintainability and reliability of modern software systems that increasingly rely on them.

\noindent\emph{Objective:}  
In this study, we investigate Software Dependencies 2.0 in open-source software (OSS) projects by examining the reuse of PTMs, with a focus on how developers manage and integrate these models.
Specifically, we seek to understand: (1) how OSS projects structure and document their PTM dependencies; (2) what stages and organizational patterns emerge in the reuse pipelines of PTMs within these projects; and (3) the interactions among PTMs and other learned components across pipeline stages. 

\noindent\emph{Method:} We conduct a mixed-methods analysis of a statistically significant random sample of \TotalProjects GitHub repositories from the PeaTMOSS dataset (28,575 repositories reusing PTMs from Hugging Face and PyTorch Hub). We quantitatively examine PTM reuse by identifying patterns and qualitatively investigate how developers integrate and manage these models in practice.

\noindent\emph{Results:} We find that multi-PTM reuse is common (\ProjectsMultiPTM of the studied projects). In these projects, a notable portion of PTMs are interchangeable (37\%), meaning one model can seamlessly replace another in the workflow, while others are complementary (23\%), performing distinct functional roles.
This integration complexity is compounded by fragmented dependency declarations across code, documentation, and configuration files, with only \ProjectsCentralizedDocumentation of projects documented outside code. We identify three types of PTM reuse pipelines, \ie feature extraction, generative, and discriminative, all involving varying degrees of PTM adaptation, from as-is reuse to head addition or architectural modification. PTMs also frequently interact with other models in tightly coupled or modular designs, reflecting the complexity of such systems.

\noindent\emph{Conclusion:} Managing Software Dependencies 2.0 and PTM reuse pipeline complexity introduces unique technical challenges in ML-enabled software.
Our findings underscore the need for enhanced tools and practices that treat PTMs as first-class, modular components, facilitating their reliable integration and long-term maintenance within Software Dependencies 2.0. 
We identify future research on PTM reuse and integration as a way to further understand Software Dependencies 2.0.

\keywords{Pre-trained Models \and Software Dependencies \and Model Reuse \and ML pipeline \and Software Supply Chain \and  Model Interactions \and Technical Debt \and Machine Learning \and Deep Learning}

\end{abstract}

\section{Introduction} \label{sec:intro}
Deep neural networks (DNNs) are driving major advances in domains such as image analysis~\citep{namysl2019efficient,yin2019deep}, natural language processing~\citep{collobert2008unified,Nakano2021WebGPTBQ,zhou2022natural}, and speech recognition~\citep{gaikwad2010review,abdel2014convolutional,nassif2019speech}.
In software engineering, DNNs are increasingly applied to tasks such as code generation and automated testing~\citep{le2020deep,ale2024enhancing,chen2025deep}.
Yet, training these models from scratch is resource-intensive, requiring substantial computational power and engineering effort as architectures grow in size and complexity~\citep{narayanan2019pipedream,geiping2023cramming}.
\textbf{\textit{Pre-trained models (PTMs)}} have emerged as a practical solution to this challenge. By encapsulating both the architecture and the parameters learned from data, PTMs enable developers to transfer knowledge across tasks, significantly reducing training costs and development time~\citep{marcelino2018transfer,han2021pre}. The recent emergence of foundation models (FMs), \ie large-scale PTMs trained on broad, heterogeneous data and adaptable to a variety of tasks~\citep{Bommasani2021}, has further accelerated PTM reuse, with models such as GPT-4~\citep{OpenAI2023} and Stable Diffusion~\citep{Rombach2022}.
Reports indicate that the global AI market, encompassing technologies like PTMs, is projected to grow from USD \$279.22 billion in 2024 to USD \$1,811.75 billion by 2030~\citep{grandview2025aimarket}.
This momentum highlights the need to better understand how PTMs are being reused in practice.


\new{Integrating PTMs as dependencies introduces several technical challenges spanning model integration, environment setup, documentation, resource configuration, and adapting PTM behavior to project-specific tasks}~\citep{jiang2024peatmoss,banyongrakkul2025release}. We believe that these problems signal a new form of software dependency, which we term \textbf{\textit{Software~Dependencies~2.0}}. We contrast this with \textit{Software Dependencies 1.0}, referring to the integration of conventional software libraries or packages built from source code (code-centric) and typically managed through mature package management ecosystems~\citep{decan2019empirical}. \textit{Software Dependencies 2.0}, by comparison, refers to the integration of software with learned behaviors encapsulated in machine learning (ML) models whose parameters are determined by data and optimization processes (model-centric). These dependencies are often distributed through model hubs as reusable artifacts and pose new challenges for the maintainability of software that incorporates them.

Recent work~\citep{jiang2024peatmoss} frames PTM development and deployment as a model supply chain, spanning upstream model producers~\citep{brown2020language,rombach2022high,OpenAI2023}, intermediary platforms~\citep{castano2023exploring,mitchell2019model}, and downstream software systems that consume these models. While prior research has primarily focused on upstream concerns such as model documentation, security, and sharing~\citep{castano2023exploring,jiang2022empirical,kathikar2023assessing,jiang2025see}, downstream reuse remains less understood. Understanding downstream PTM reuse is crucial: without it, systems may become harder to maintain and reproduce~\citep{decan2019empirical,okafor2022sok}, and upstream producers may miss key requirements for designing PTMs, platforms, and tools. From the perspective of \textit{Software Dependencies 2.0}, understanding downstream practices reveals how model-centric dependencies are managed and integrated into software.

In this paper, \textbf{we use PTMs as a lens to study \textit{Software~Dependencies~2.0}, focusing on their management and integration into ML-enabled systems, including the interaction between multiple models within such systems}. 
By integration, we refer to the entire process through which PTMs are incorporated into software projects, encompassing multiple stages such as initialization of models and adapting them to project-specific requirements. Understanding these stages provides a structured view of PTM reuse and informs our analysis of practical challenges and patterns. Our study is organized around three research questions:

\begin{enumerate}[label=\textbf{RQ\textsubscript{\arabic*:}}, leftmargin=4.5em]
    \item \rqone
    \item \rqtwo
    \item \rqthree
\end{enumerate}

To address these questions, we use the PeaTMOSS dataset~\citep{jiang2024peatmoss}, which contains 28,575 GitHub repositories that employ models from Hugging Face and PyTorch Hub. From this dataset, we extract a stratified statistically representative sample of 650 popular projects (341 using PTMs from Hugging Face Hub and 314 from PyTorch Hub). After categorizing these projects by the functional role of PTMs, we focus our analysis on a manually-curated set of 401 projects that use PTMs to implement core functionality. 

We adopt a mixed-methods approach to analyze these target projects. For RQ1, we detect and trace PTM usage across these projects, examining how their dependencies on PTMs are structured and documented. For RQ2 and RQ3, we first review literature from the SE and ML domains to establish a theoretical grounding on PTM reuse pipelines and their interactions with other learned components (\ie PTMs or scratch-trained models) in ML-enabled systems. We then analyze the same 401 projects to qualitatively characterize real-world reuse and multi-model interaction patterns, comparing our empirical observations with theoretical expectations.

\new{The main findings from our RQs are:}
\begin{itemize}[leftmargin=*, itemsep=2pt, topsep=2pt, parsep=0pt, partopsep=0pt]
\item \new{\textbf{RQ1:} We observed various practices in structuring and declaring dependencies on PTMs (§4.1). Among multi-PTM projects, dependencies are either interchangeable (37\%) or complementary (23\%). Dependencies are often inconsistently specified across source code, configuration files, and README files, with limited version control.}

\new{\item \textbf{RQ2:} We analyzed the stages of PTM reuse and identified ten recurring stages (§4.2). Projects commonly organize these stages into three pipeline types: feature extraction, generative, and discriminative. Across pipelines, PTMs are reused either as-is or with adaptation (e.g., head addition or architectural modification), with adaptation being far more common, suggesting that PTM reuse is rarely plug-and-play. Compared to conventional ML pipelines~\citep{amershi2019software,biswas2022art}, PTM-centric pipelines often bypass or restructure standard stages such as model training or feature engineering, introducing new maintenance challenges.}

\new{\item \textbf{RQ3:} PTMs frequently interact with other models (in 50\% of the studied projects). 
We developed a taxonomy of four recurring interaction types (§4.3): feature 
handoff, feedback guidance, evaluation, and post-processing refinement. }

\end{itemize}

\new{To better capture the emerging characteristics observed in our empirical study, we introduce the concept of \textit{Software Dependencies 2.0}, formally defined below.}

\FindingsBox{\new{Definition: Software Dependencies 2.0}}{
\new{
Synthesizing the results of our empirical study, we define \emph{Software Dependencies 2.0} as model-centric dependencies whose behavior is determined by learned parameters rather than explicit source code, and whose integration requires managing semantic, pipeline-level, and cross-model interactions.
This definition refines our earlier intuitive framing of Software Dependencies 2.0 as the integration of software with learned behaviors encapsulated in ML models.
Unlike Software Dependencies 1.0 (code-centric), which are typically declared through static manifests and integrated via stable APIs, Software Dependencies 2.0 introduce dependencies whose meaning emerges from reuse pipelines, adaptation strategies, runtime configuration, and interactions with other models.
As summarized in Table~\ref{tab:dependencies_2}, these dependencies differ from conventional ones along many dimensions:
\begin{itemize}[leftmargin=*]
\item Dependency structure shifts from static edges (``depends on'') to semantic and cross-stage interactions.
\item Integration involves both code APIs and model artifacts, \eg pretrained weights, tokenizers, and adapters, often requiring adaptation such as fine-tuning or head selection.
\item Interchangeability and complementarity emerge at the model-family and pipeline level rather than at rigid API boundaries.
\item Behavior is context-sensitive and probabilistic rather than deterministic.
\item Traceability is often decentralized and implicit, \ie scattered across code, configuration files, and runtime context, rather than centrally declared in manifests.
\end{itemize}
PTM reuse provides a canonical instantiation of Software Dependencies 2.0, enabling us to ground this definition in observed practices across real-world systems.
}
}
In summary, our study provides the following main contributions to understanding Software~Dependencies~2.0:

\begin{itemize} 
    \item We provide the first systematic analysis of how PTMs are defined and managed as dependencies in OSS projects, highlighting widespread inconsistency and fragmentation (\S\ref{subsec:rq1}).
    \item We characterize PTM reuse pipelines by identifying three dominant orientations and common adaptation strategies (\S\ref{subsec:rq2}).
    
    \item We propose a taxonomy of four multi-model PTM interaction types, revealing additional architectural complexity (\S\ref{subsec:rq3}).

    \item We conceptualize Software~Dependencies~2.0, highlight key engineering risks related to PTM reuse, and clarify major differences between Software Dependencies~1.0 and~2.0, motivating the need for improved tooling and further research. We also provide suggestions for different stakeholders involved in PTM integration (\S\ref{sec:discussion}).
\end{itemize}

\noindent \textbf{Significance:}
This study is the first to characterize the reuse patterns of PTMs sourced from public model hubs in real-world software projects, examining them as a new kind of software dependency that differs from conventional library reuse. Our findings highlight technical challenges that have been largely overlooked in the SE literature on PTM reuse. At the same time, our study complements the ML literature by bridging the gap between model design and real-world system integration.

\vspace{0.1cm}
\noindent \textbf{Paper organization:}
Section \ref{sec:background} discusses the background on software dependencies, PTM, and PTM reuse. Section \ref{sec:data_processing} motivates the RQs and presents the study design. Section \ref{sec:empirical_inv} details the methodology and results for RQs. Section \ref{sec:discussion} presents discussions and implications of our findings. Section \ref{sec:threats} discusses threats to validity, Section~\ref{sec:related_work} discusses related work, and Section \ref{sec:conclusion} concludes.
We describe our reproducibility artifact in Section~\ref{sec:declarations}.

\section{Background}\label{sec:background}
Reusing pre-trained models (PTMs) has become a common practice in modern software systems. Such reuse introduces a new type of software dependency that differs from conventional libraries. In this section, we provide background on three key topics. First, we review \textit{Software Dependencies}. Second, we discuss PTMs as ``Software~Dependencies~2.0'', highlighting their unique characteristics. Third, we describe \textit{PTM Reuse Pipelines} and \textit{cross-stage multi-model interactions}, focusing on how multiple models are composed and coordinated within complex systems.

\subsection{Software Dependencies}
\label{subsec:software_dep}
Software systems are rarely built in isolation; they integrate external components that provide functionality through well-defined interfaces~\citep{sonatype_sscr2024}.
These relationships are commonly referred to as \textit{software dependencies}. A software dependency represents a relationship in which one software component relies on another to function correctly. This dependence can occur through function calls, data access, or library usage. We refer to the typical dependencies with which the reader is familiar --- logging libraries, web frameworks, wrappers to interact with different services --- as \textit{Software Dependencies 1.0}. By this we mean the \underline{code-centric} components whose behavior is explicitly programmed. In widely used package ecosystems such as PyPI and NPM, dependencies are typically declared in manifest files (\eg \texttt{requirements.txt}, \texttt{package.json}) and resolved automatically by package managers~\citep{wittern2016look, decan2019empirical}. 
In this context, dependency management primarily concerns versioning, API stability, and security risks, rather than the behavior of the dependency itself, since the functionality of these components is fully defined by human-written code~\citep{pashchenko2020qualitative}. While most dependencies require little to no configuration, there are occasional cases of components with more extensive configurability~\citep{clements2002software}.

In contrast, the reuse of PTMs in downstream projects introduces a new class of software dependency, which we term \textit{Software Dependencies 2.0}. These components extend beyond human-written code to include reusable, \uline{model-\allowbreak centric} artifacts such as model architecture, pretrained parameters, vocabulary files, and configuration settings~\citep{jiang2022empirical,castano2023model}, all of which are essential for correct usage and integration in downstream systems.  PTMs are commonly distributed as installable packages through centralized model registries, which facilitate adoption, versioning, and integration. This distribution mechanism enables downstream applications to leverage PTMs’ learned behaviors much like libraries, but with the additional complexity introduced by the model-centric nature of the artifacts.~\citet{jiang2023empirical} previously observed that packages differ structurally and functionally; our study, however, focuses on the implications of these dependencies for integration and management in downstream systems. Recent qualitative evidence from \citet{jiang2025see} on PTM package development further suggests that these properties lead to different integration and management practices, which we investigate in this study.




\subsection{Pre-Trained Models (PTMs) and Software Dependencies 2.0} 
\label{subsec:ptm_dependency}

PTMs represent neural network architectures that have undergone prior training in large datasets, enabling them to generate output or adapt for downstream tasks~\citep{gonzalez2025pre}. These models encapsulate learned parameters, often including a defined data pipeline and training regime, making them distinct from conventional software components whose behavior is explicitly programmed~\citep{karpathy2017software2}.

PTMs are commonly distributed as \textit{PTM packages}, which bundle the model architecture, pre-trained weights, metadata such as training datasets, data preprocessing steps, and evaluation metrics~\citep{jiang2024peatmoss}. These packages are typically shared through \textit{model registries} or \textit{hubs}, which facilitate adoption and fine-tuning of PTMs for specialized applications~\citep{jiang2024peatmoss}.
Examples include open registries like Hugging Face~\citep{hugging2021} and PyTorch Hub~\citep{torchhub}, and commercial registries such as NVIDIA NGC~\citep{NVIDIANGC} and the Qualcomm AI Hub~\citep{Qualcomm}.

The scale of these registries has grown rapidly; for example, Hugging Face’s model hub expanded from just a few thousand models in 2022 to over one million models by 2024~\citep{huggingface2024review}, reflecting the accelerating community adoption. At the same time, usage has intensified: downloads of the top 10\% of PTMs nearly doubled within a single year, growing from 269 billion in August 2022 to 587 billion in August 2023~\citep{jiang2024peatmoss}.

\begin{figure}[!htb]
    \centering
    \fbox{
        \includegraphics[width=0.98\linewidth]{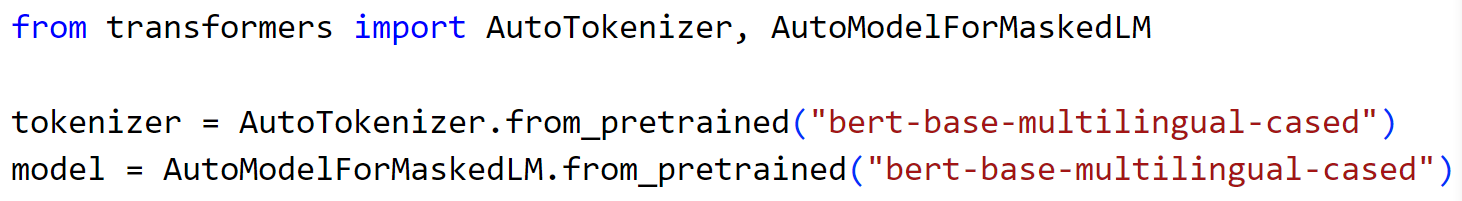}
    }
    \caption{
     Example of PTM reuse in the Hugging Face ecosystem.
     This code snippet loads a multilingual BERT model and tokenizer using the \code{transformers} library. While reuse appears straightforward at the API level, the underlying model introduces complex dependencies, including learned behavior, format constraints, and resource requirements, that are not immediately visible in code. Such \textit{Software Dependencies 2.0} represent a distinct and underexplored dimension of software reuse. Beyond dependencies, PTM reuse also involves broader aspects of integration and interaction between models, highlighting the complexity of real-world usage.
    }
    \label{fig:usage_HF}
\end{figure}

Downstream applications adopt PTMs as \textit{dependencies} to leverage pretrained capabilities. For example, in the Hugging Face ecosystem, developers can load a tokenizer and a multilingual BERT model via the \code{transformers} library, as shown in Figure~\ref{fig:usage_HF}.

Managing Software Dependencies 2.0 involves familiar concerns such as versioning \citep{ajibode2025towards} and compatibility, but the probabilistic and data-driven nature of PTMs adds complexity. Updates or retraining can unpredictably alter model behavior, complicating validation and increasing the risk of subtle errors \citep{jiang2022empirical}. These challenges are further amplified by the computationally intensive and specialized runtime environments often required by PTMs \citep{guo2022threats,banyongrakkul2025release}.

The multiple interdependent components of a PTM, such as its architecture, learned parameters, and auxiliary assets, further complicate integration and introduce risks rarely encountered with conventional libraries. Understanding these parallels and divergences is critical for developing dependency management practices that address both inherited challenges and novel risks introduced by PTMs. While prior research has documented difficulties in conventional ML systems \citep{sculley2015hidden,bogner2021characterizing}, the practical challenges of integrating and maintaining PTMs in real-world projects remain less explored, motivating a systematic investigation of PTM reuse practices. In this work, we systematically investigate PTM reuse and integration in open-source projects, focusing on model-centric dependencies.

\subsection{PTM Reuse Pipelines and Cross-stage Multi-model Interactions}
\label{subsec:ptm-stages}

\begin{figure*}[!ht]
\centering
\includegraphics[width=0.99\linewidth]{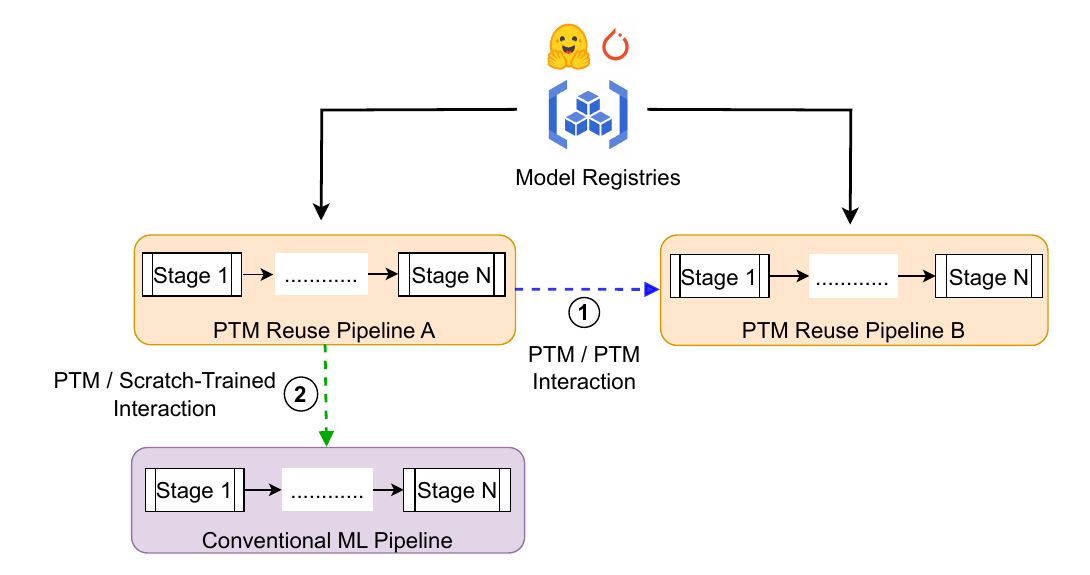}
\caption{Overview of PTM reuse pipelines and their interactions within a project.
PTM Reuse Pipeline~\textit{A} reuses models from registries such as Hugging Face and PyTorch. It may interact with other pipelines in two ways:
PTM/PTM Interaction (\ding{172}): interaction with PTM Reuse Pipeline~\textit{B}.
PTM/Scratch-Trained Interaction (\ding{173}): interaction with Conventional ML Pipeline.
Dashed arrows indicate optional steps.
Ellipses show omitted intermediate stages.
}
\label{fig:background}
\end{figure*}

\new{To characterize how PTMs are reused in practice, we introduce the notion of \textit{PTM reuse pipelines}, i.e., the operational structures through which PTM dependencies are integrated, configured, and composed within software systems.} A \textit{PTM reuse pipeline} is a structured sequence of stages through which a pretrained model progresses from acquisition to operational use.
\new{Fig.~\ref{fig:background} presents a conceptual framework for understanding PTM reuse pipelines, drawn from established literature on software architecture and ML workflows~\citep{amershi2019software, biswas2022art}. This background framework provides the conceptual foundation for our analysis of real-world PTM usage patterns in Section \ref{sec:empirical_inv}.}

\begin{figure*}[t]
\centering
\includegraphics[width=0.99\linewidth]{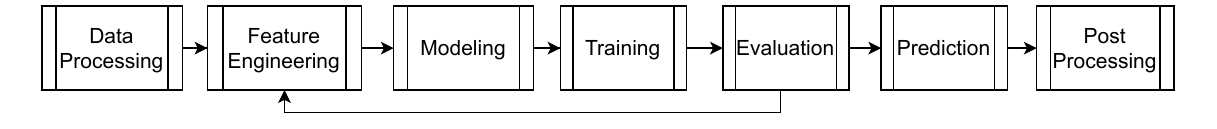}
\caption{Overview of conventional ML pipelines. Stages such as data processing and post-processing are treated at a coarse level, as their fine-grained substeps (\eg data acquisition, data preparation) vary widely and are often implicit or domain-specific.}
\label{fig:background_conv_pipeline}
\end{figure*}

The notion of a PTM reuse pipeline builds on the broader concept of software and data processing pipelines, introduced by~\citet{mary1996software} as an architectural style in which components (filters) are connected by data flows (pipes). 
~\citet{garlan2000software} further developed this idea with descriptions and examples that help software developers design and communicate complex systems in a clear and intelligible manner.
In conventional ML~\footnote{Conventional ML refers to models trained from scratch on data.}, pipelines are the dominant abstraction for workflows involving data processing, modeling, training, and post-processing~\citep{amershi2019software, biswas2022art}, as shown in Fig.~\ref{fig:background_conv_pipeline}.
We extend this convention by treating the PTM reuse pipeline as a reusable architectural unit distinct from both the PTM itself and the broader software system.

PTM reuse can involve multiple models within a project, including PTMs and scratch-trained models, resulting in diverse reuse pipelines that may interact with one another. We refer to such interactions as \textit{cross-stage multi-model interactions}, \ie cases where the output of one pipeline (features, representations, or predictions) serves as input, context, or augmentation for another.
Research on conventional ML model integration has shown that systems often contain interconnected models organized into functional configurations~\citep{peng2020first, sens2024large}. For example,~\citet{peng2020first} analyze ML architectures in autonomous driving systems and reveal patterns of model interconnection within monolithic applications, though their work focuses on scratch-trained components and does not specifically address PTM reuse. Similarly,~\citet{sens2024large} provide a broader survey across 23 real-world projects, identifying multi-model configurations such as alternative models (for the same task), independent models (serving different tasks), as well as sequential chains, and joining integrations of models. While valuable, these classifications primarily reflect co-existence rather than interaction. Even sequential and joining patterns are described at a coarse system level, without attention to the structured reuse of PTMs or pipeline-based composition. In other words, prior work either analyzed code-level interactions of scratch-trained models or provided high-level system-wide patterns of model interactions, but did not examine how PTMs and optional scratch-trained models interact across stages within reusable pipelines.

In contrast, our work centers on how PTMs are reused through pipelines and how the pipelines of PTMs and scratch-trained models interact with one another across different stages. Each model—whether pretrained or scratch-trained—can be associated with its own operational pipeline. We therefore analyze interactions not solely at the model level, but in terms of how these pipelines exchange data, features, or representations across stages. In this framing, one of the interacting pipelines is always a PTM reuse pipeline, while the other may be built around a scratch-trained model or another reused model. This approach offers a more nuanced understanding of compositional integration and modularity compared to existing taxonomies. To facilitate our analysis of PTM reuse in these settings, we identify two categories of cross-stage multi-model interactions, grounded in patterns observed across the studied projects:

\begin{enumerate}
\item \textbf{PTM/PTM interactions}: 
    Interactions in which two or more PTM reuse pipelines within the same project coordinate to fulfill complementary roles. Figure~\ref{fig:background} (\ding{172}) illustrates this with PTM reuse Pipeline~A and PTM reuse Pipeline~B, each using models from registries such as Hugging Face or PyTorch.  

\item \textbf{PTM/Scratch-trained interactions}: 
    Interactions in which a PTM reuse pipeline and a scratch-trained model pipeline coordinate to enhance, support, or augment each other within the same project. Figure~\ref{fig:background} (\ding{173}) illustrates this interaction between PTM reuse Pipeline~A and the conventional ML (scratch-trained) pipeline.
\end{enumerate}

\section{Study Design}\label{sec:data_processing}
This section describes the design of our empirical study. We first present our research questions and study overview (Section~\ref{subsec:study_overview}), then detail the shared data preparation steps that underpin all three research questions (Section~\ref{subsec:data_collection}). These steps produce a dataset suitable for downstream analyses. While all RQs draw from this common foundation, each applies additional filtering or annotation in Section~\ref{sec:empirical_inv} to address its specific focus.

\subsection{Research Questions and Study Overview}
\label{subsec:study_overview}

The goal of this study is to conduct the first empirical investigation of Software Dependencies 2.0 in open-source software projects, focusing on PTMs accessed via model hubs. We examine how PTM dependencies are structured and documented (RQ1), how PTMs are organized within reuse pipelines (RQ2), and how they interact with other learned components across stages (RQ3). While RQ1 directly addresses dependency specification, RQ2 and RQ3 focus on pipeline organization and model interactions, thereby providing a more comprehensive view of how these dependencies are used and integrated in practice.
Through this investigation, we aim to uncover characteristics of Software Dependencies 2.0, identify real-world PTM reuse patterns, highlight associated software engineering risks, and inform the development of improved tooling and practices.

Our research is guided by three questions:
\begin{enumerate}[label=\textbf{RQ\textsubscript{\arabic*:}}, leftmargin=4.5em]
    \item \rqone
    \new{ To structure our analysis, we further decompose this research question into the following sub-research questions:
    \begin{itemize}
    \item \textbf{RQ1.1:} How do OSS projects structure their PTM dependencies?
    \item \textbf{RQ1.2:} Where and how are PTM dependencies documented in OSS projects?
    \end{itemize}}
    \item \rqtwo
    \new{To structure our analysis, we further decompose this research question into the following sub-research questions:
    \begin{itemize}
    \item \textbf{RQ2.1:} Which stages characterize real-world PTM reuse pipelines, and how do they differ from conventional ML pipeline stages?
    \item \textbf{RQ2.2:} How are these stages organized within OSS projects?
    \end{itemize}}
    \item \rqthree
\end{enumerate}

We begin with \textbf{RQ1}, which focuses on the structure and documentation of PTM dependencies. Conventional libraries are typically declared in configuration or manifest files (e.g., package.json, pom.xml) that are processed by package managers to install and manage them. PTMs, by contrast, combine executable code with learned artifacts~\citep{jiang2023empirical} and are often integrated through local model checkpoints or inline hub references~\citep{jiang2024peatmoss}. 
\new{Based on prior observations, there is currently no widely adopted standardized package manager for PTM dependencies.~\citep{jiang2024peatmoss}.}
As a result, it is difficult to trace which PTMs are reused within a project and how they are reused. This traceability forms the critical foundation for reasoning about their reuse and integration, which we explore further in RQ2 and RQ3. RQ1, therefore, takes the first step by asking: How many PTMs are reused within a project? How do these PTMs relate to one another (e.g., are they interchangeable or complementary)? And where and how are these dependencies documented? By addressing these questions, RQ1 establishes an understanding of how PTM dependencies are structured and documented in practice, providing the basis for deeper analysis in subsequent RQs.

\textbf{RQ2} examines how PTM reuse pipelines diverge from conventional ML pipeline structures (see Figure~\ref{fig:background_conv_pipeline}), which typically consist of standardized stages, including data preparation, feature engineering, modeling, training, and evaluation.
\new{RQ2 focuses specifically on PTM reuse, examining how PTMs are incorporated, adapted, and operationalized within PTM reuse pipelines.}

Because PTMs are pre-trained and reusable, PTM reuse may lead to the introduction of alternative or additional stages, such as inference, fine-tuning, or model adaptation. Characterizing these real-world reuse pipelines is essential for understanding how developers adapt and integrate PTM dependencies into their systems, revealing practical reuse patterns and the underlying complexity of integration.

\textbf{RQ3} explores how multiple models interact within a single project. In practice, developers can connect models either by chaining them sequentially (where the output of one model feeds into another) or by combining them in parallel~\citep{sens2024large}. 
PTMs, due to their modularity and adaptability, can support cross-stage interactions (\S\ref{subsec:ptm-stages}) by serving diverse roles across pipelines.
Studying these interaction patterns provides critical insights into modular system composition, integration complexity, and robustness.

\begin{figure*}[t]
\centering
\adjustbox{cfbox=black 1pt 3pt}{
    \includegraphics[width=0.93\linewidth]{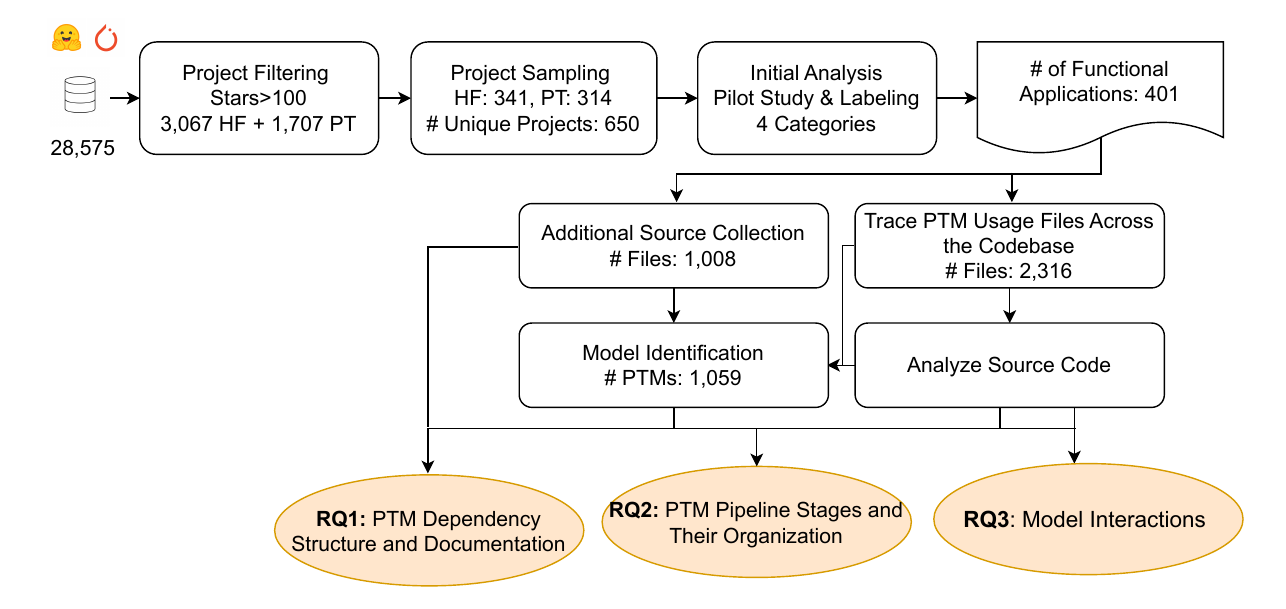}
}
\caption{\new{Data Preparation: Projects were filtered and analyzed to investigate PTM usage, pipeline stages, and model interactions.}
 }
\label{fig:study_design}
\end{figure*}

To \new{answer} these research questions, we systematically extract and analyze PTM-related artifacts such as model loading scripts, configuration files, and downstream usage code from OSS repositories. Our analysis is further enriched with insights from hub documentation and existing integration guides. 
Figure~\ref{fig:study_design} illustrates our methodological framework. 


\subsection{\new{Data Preparation}} \label{subsec:data_collection}
As illustrated in Figure~\ref{fig:study_design}, we collected candidate projects that use PTMs from two major model hubs (Hugging Face and PyTorch Hub), applied filtering and stratified sampling, and then further refined the selection based on their PTM reuse category to identify relevant targets for addressing our three research questions. In the rest of this subsection, we first detail the data collection and sampling process, followed by our approach for tracing PTM usage within the selected repositories.

\vspace{0.5em}
\noindent\textbf{Data Source.}
To study PTM reuse in real-world software, we need a dataset of repositories that actually incorporate pretrained models. We examined the available datasets~\citep{jones2024we} and identified the PeaTMOSS dataset \citep{jiang2024peatmoss} as suitable for our purpose.
The PeaTMOSS dataset~\citep{jiang2024peatmoss} contains 28,575 GitHub repositories that reuse PTMs from Hugging Face and PyTorch Hub. These repositories were identified through a signature-based approach that targets project source code, where a signature is defined as the combination of a specific library import (\eg \texttt{transformers}) and a corresponding function call used to load a PTM (\eg \texttt{from\_pretrained}).

The dataset captures repositories available as of July 10, 2023. For our analysis, we collected the February 2024 snapshot of these repositories.

\vspace{0.5em}

\noindent\textbf{Project Filtering and Sampling.}
Analyzing all 28,575 projects is infeasible due to the intensive manual effort required to examine source code and answer our three research questions. Therefore, we curate a representative subset of projects through a multi-step filtering process.

First, we filter by project popularity, including only repositories with at least 100 GitHub stars, an approach consistent with prior work~\citep{han2019characterization}, to focus on impactful OSS projects. This yields 3,067 repositories that reuse models from Hugging Face and 1,707 repositories that reuse models from PyTorch Hub.
Next, to ensure statistical representativeness while keeping the analysis tractable, we apply stratified random sampling by hub (Hugging Face and PyTorch Hub) with a 95\% confidence level and a 5\% margin of error. Stratifying by hub was chosen because the two hubs differ in their typical domains (\eg NLP, computer vision, audio). This results in a final sample of 341 Hugging Face and 314 PyTorch Hub repositories. \new{Five repositories appearing in both hubs were counted once in the total number of projects, resulting in 650 unique projects.}

\vspace{0.5em}
\noindent\textbf{Initial Analysis.} Not every software project that uses a PTM qualifies as a target PTM reuse project for our study, as some may simply host tutorials demonstrating PTM usage. To better understand the nature of PTM usage, we conducted an initial analysis of how the \PreliminaryTotalProjects sampled projects incorporate PTMs.

We adopt an open coding approach~\citep{seaman1999qualitative} to manually review each repository. For simplicity, rather than analyzing the full source code, we examined only the README files, which typically summarize project goals, usage, and functionality in a concise and accessible manner.

\begin{itemize}
    \item \textbf{Pilot Construction:} 
    A random sample of 30\% of the projects that reuse PTMs from each hub 
    (105 from Hugging Face and 95 from PyTorch) was selected for manual review. Two authors collaborated to develop an initial set of categories that reflect the diversity of projects across both hubs.\footnote{Both authors have over 9 years of software engineering experience, including 5 years working with pre-trained models, ensuring accurate project categorization.}

    \item \textbf{Extension to the Remaining Projects:} 
    The same two authors annotated the remaining projects based on the initial categories, achieving a Cohen's kappa coefficient of 0.79 for Hugging Face and 0.80 for PyTorch, indicating substantial inter-rater agreement~\citep{viera2005understanding}. \new{
Disagreements were resolved through careful discussions to ensure consistent interpretation of category boundaries. Because our categorization targets high-level PTM reuse roles rather than fine-grained implementation details, the categories were intentionally coarse-grained and stabilized during the pilot phase. No additional categories emerged during the extension phase, and all disagreements were resolved through consensus.}
\end{itemize}

Our categorization reveals four primary types of PTM reuse, as summarized in Table~\ref{tab:reuse_categories}. Among the sampled projects, 61.5\% are classified as Functional Applications, followed by 27\% as PTM Facilitation Frameworks. Educational Resources comprise 11\%, and 0.05\% falls under the Other category.

\newcolumntype{P}[1]{>{\raggedright\arraybackslash}p{#1}}

\begin{table}[htbp]
\centering
\begin{threeparttable}
\caption{
Categories of PTM reuse in OSS projects. Our study is focused on Functional Applications.
}
\label{tab:reuse_categories}
\begin{tabular}{@{} P{2.5cm} P{4.4cm} P{3.3cm} @{}}
\toprule
\textbf{Category} & \textbf{Definition} & \textbf{Example Projects} \\
\midrule
Functional applications & 
Projects that employ PTMs to implement core functionalities such as question answering, image generation, object recognition, text classification, or other domain-specific tasks. & 
\texttt{AudioLDM}\tnote{1} \ (generates speech and sounds), \texttt{CodeRL}\tnote{2} \ (code generation and assistance) \\
\addlinespace[0.7ex]

PTM facilitation frameworks & 
Projects that provide tools, APIs, or libraries to support the reuse, fine-tuning, and deployment of PTMs. These typically focus on enabling broader and easier access to PTMs. & 
\texttt{Quaterion}\tnote{3} \ (similarity learning framework), \texttt{Espresso}\tnote{4} \ (toolkit for ASR with GPU training) \\
\addlinespace[0.7ex]

Educational resources & 
Projects that aim to educate developers on PTM usage, offering tutorials, documentation, and example implefmentations. & 
\texttt{Hugging Face Notebooks}\tnote{5} \ (research code and tutorials) \\
\addlinespace[0.7ex]

Other & 
Projects that replicate prior work, offer benchmarks, or provide datasets and metrics for PTM evaluation. & 
\texttt{MLCommons}\tnote{6} \ (reference implementations for MLPerf benchmarks) \\
\bottomrule
\end{tabular}

\begin{tablenotes}
\footnotesize
\item[1] See \url{https://github.com/haoheliu/AudioLDM}.
\item[2] See \url{https://github.com/microsoft/CodeRL}.
\item[3] See \url{https://github.com/UKPLab/sentence-transformers}.
\item[4] See \url{https://github.com/espnet/espnet}.
\item[5] See \url{https://github.com/huggingface/notebooks}.
\item[6] See \url{https://github.com/mlcommons/inference}.
\end{tablenotes}

\end{threeparttable}
\end{table}

\textit{Given that our study focuses on understanding how PTMs are integrated into downstream applications, we limit our in-depth analysis to the \textbf{401} projects categorized as Functional Applications.}
Among the 401 Functional Application projects, star counts range from 100 to 4,200, with a median of 320 and an average of 540~\citep{borges2018s,he20244}. 

\vspace{0.5em}  

\noindent\textbf{Tracing PTM Usage Files Across the Codebase.}
To identify the subset of files transitively involved in PTM usage, we develop a lightweight tracing procedure. The PeaTMOSS dataset provides method signatures for identifying PTM-loading functions (\eg \texttt{from\_pretrained()}) via static analysis, but it only locates model-loading points. Analyzing every file in each repository is infeasible due to their size and complexity. Our procedure focuses on files that participate in the PTM reuse pipeline, a structured sequence of stages from model acquisition to operational use (\S\ref{subsec:ptm-stages}). We begin by locating model-loading calls using PeaTMOSS signatures and \texttt{git grep} to search for matching patterns. For each match, we identify the enclosing function or method (\eg \texttt{load\_model}) using static analysis.

These functions serve as the starting points for our tracing procedure. We then perform an iterative, breadth-first traversal of the call graph using \texttt{git grep}, identifying where each function is called and extracting the enclosing caller functions. We set the depth limit to 6 to balance completeness and noise. Based on a pilot study of \new{120} randomly sampled projects \new{(approximately 30\% of the dataset)}, we observed that more than 95\% of PTM-related activities occur within 3 to 4 levels of indirection from the model load call. Beyond 6 levels, the likelihood of encountering unrelated or utility function references increases significantly.

Since our tracing is based on static patterns, it may miss dynamic method calls (\eg reflection, dynamic dispatch) or implicit data flows (\eg via global state or object attributes). To mitigate these limitations, we manually inspect cases where tracing appears incomplete or ambiguous, especially in projects with non-standard control flow. For 15 projects, we have identified the relevant files during this manual inspection. This manual step helps validate results and recover connections that static analysis alone may miss.

\vspace{0.5em}

\noindent\textbf{Additional Source Collection.} In addition to PTM-usage files, we identify several other types of files for further analysis. Our collection includes \texttt{README} files, which provide insight into the project. Configuration files are collected as they define the settings and parameters necessary for model usage, including architecture configurations, training hyperparameters, and optimization techniques. We collect all configuration files with~\texttt{.yaml},~\texttt{.yml},~\texttt{.xml},~\texttt{.config}, or~\texttt{.json} extensions.\footnote{In our dataset, 95\% of projects that included configuration files used these extensions.}

\vspace{0.5em}
\noindent \new{\textbf{PTM Name Identification.} 
We identify the specific PTMs used in each project by extracting model identifiers from loading statements. PeaTMOSS provides direct mappings between known loading functions and PTM names when those names are \textit{statically defined}, \ie hardcoded string literals in the source code. These mappings are already included in the dataset. For example:}
\new{
\begin{itemize}
    \item \texttt{AutoModel.from\_pretrained("bert-base-uncased")} \hfill (HuggingFace Transformers)
    \item \texttt{torchvision.models.resnet18(pretrained=True)} \hfill (TorchVision)
\end{itemize}
}
\new{
In contrast, for \textit{dynamically loaded} models, names are not fixed in the code but stored in variables, passed as arguments, or read from configuration files or command-line inputs. Although PeaTMOSS includes the loading function (\eg \texttt{from\_pretrained(...)}), it does not identify the exact model name. We start from the files where these calls appear and manually check nearby code and other project files collected in previous steps to trace the data flow and identify the actual model names being loaded. }

\begin{figure}[htb]
    \centering
    \adjustbox{cfbox=black 1pt 3pt}{
    \includegraphics[width=0.8\linewidth]{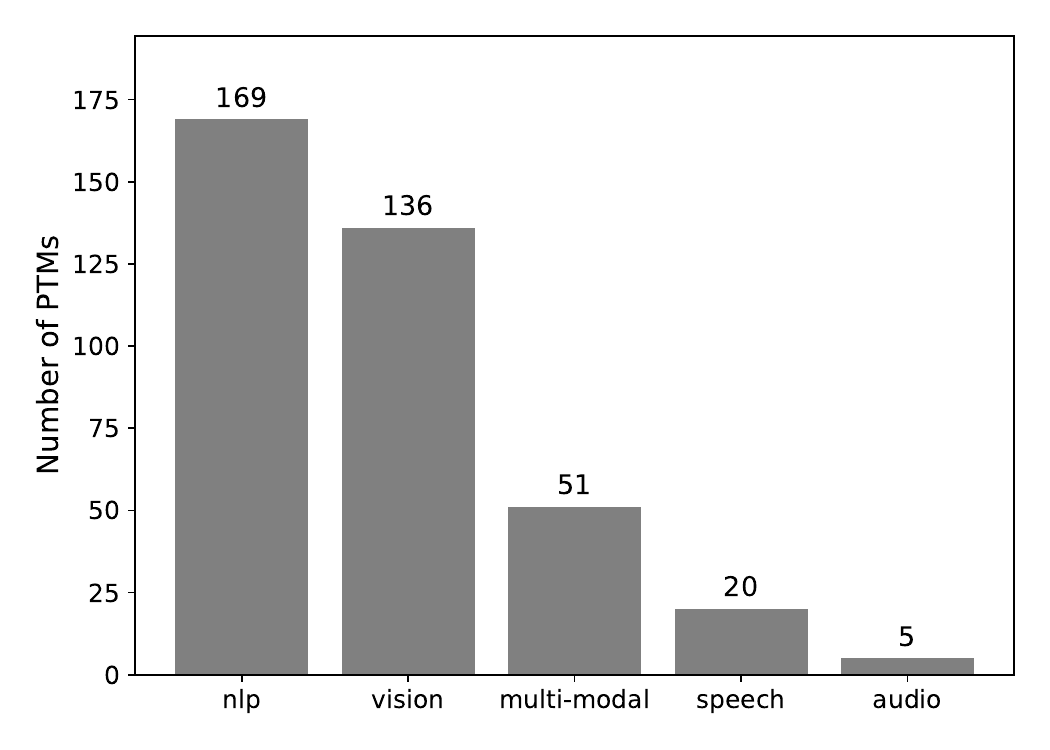}}
    \caption{\new{Number of unique PTMs by modality in the target 401 projects.}}
    \label{fig:model_types}
\end{figure}

\new{
\subsection{Statistics of Identified PTMs}
\label{subsec:ptm-stats}}
\new{
In total, we identify 1,059 PTM usages across the 401 projects, involving 381 unique models. Figure~\ref{fig:model_types} shows the distribution of the 381 unique models by modality (as provided by PeaTMOSS). NLP models are the most diverse (169 models, 44.4\%), followed by vision (136, 35.7\%), multi-modal (51, 13.4\%), speech (20, 5.2\%), and audio (5, 1.3\%). The models span from 2012 to 2024, with the largest concentration in 2019--2022 (63.7\%). Architecturally, Transformer-based models make up 61.1\%, including encoder-only (\eg \texttt{BERT}, \texttt{ViT}; 36.6\%), decoder-only (\eg \texttt{GPT-2}, \texttt{LLaMA}; 15.3\%) and
encoder--decoder (\eg \texttt{T5}, \texttt{BART}; 9.2\%).}

\vspace{0.5em}

\section{Detailed Methods and Results}\label{sec:empirical_inv}
This section presents the detailed method and results for each of our RQs. 

\subsection{\textbf{RQ1: \rqone}}
\label{subsec:rq1}

\noindent To answer RQ1, we manually analyze both the \textit{structure} of dependencies and their \textit{documentation}. This analysis leverages \TotalProjects projects processed as described in \S\ref{subsec:data_collection}.

\subsubsection{Methodology}
\label{subsec:rq1_methodology}

\paragraph{Step 1: Manual Examination of PTM Dependency Structure (for RQ1.1).}

Using the PTM names extracted in~\ref{subsec:data_collection}, we first classify each repository as using either a single PTM or multiple PTMs. For repositories containing multiple PTMs, we examine the relationships between them at the project level, specifically asking whether the PTMs used within a project belong to the same base family or span across different families. 
To support this analysis, we group individual PTMs into broader \textit{base model families}.
\new{We define a base model family as a group of pretrained models that share the same underlying neural architecture and primary pretraining objective, while differing in training data, scale, or optimization strategy. This definition captures models that are functionally similar in their learned representations despite variations in training methodology. For instance, the BERT family includes \texttt{bert-base-uncased}, \texttt{roberta-base}, \texttt{distilbert-base-uncased}, and \texttt{albert-base-v2}. Compared to the original BERT model, RoBERTa modifies the optimization and training configuration, \texttt{DistilBERT} reduces model size via distillation, and \texttt{ALBERT} introduces parameter sharing. Despite these differences, all retain the bidirectional transformer encoder with masked language modeling pre-training and are therefore considered part of the same base model family. This categorization aligns with the existing literature that treats models such as \texttt{RoBERTa} and \texttt{DistilBERT} as \texttt{BERT} variants~\citep{liu2019roberta,sanh2019distilbert}.}

We assign models to (base model) families using a two-step process. First, we apply a keyword-based heuristic that relies on architecture-identifying substrings in model names. For instance, any model whose name contains ``bert'' was assigned to the \texttt{BERT} family, ``gpt'' to \texttt{GPT}. This approach is supported by prior work~\citep{jiang2023naming,hao2023mgit}, which reports that Hugging Face model names generally reflect underlying architectures. Second, when model names do not explicitly include the family name, we manually map them to the appropriate family. For example, inpainting models such as \texttt{anything-4.0-inpainting}, which omit ``Stable Diffusion'' but are known community variants,  \texttt{Stable Diffusion} family based on official documentation~\citep{hugging2021,torchhub}, and our prior knowledge. 
\new{One author performed the initial labeling, and a second author independently reviewed all labels as a validation step. The percent agreement between the reviewer and the original annotations was 85\%. All discrepancies were discussed and resolved.} Note that Hugging Face provides metadata fields such as \texttt{model\_type} for each model hosted on the hub. However, this metadata does not consistently capture shared architectures and is therefore unreliable for identifying model families~\citep {jiang2025see}.

In addition, for projects containing multiple PTMs, we inspect the source code to understand how these models are integrated. Projects are labeled as interchangeable if they included PTMs that could be swapped within the workflow without altering the surrounding pipeline, such as models appearing in optional lists or conditional branches (\eg \texttt{if-else} blocks). Otherwise, they are labeled as complementary. Some projects exhibited both types.

\paragraph{Step 2: Identify PTM Name Declarations (for RQ1.2).} Building on the PTMs identified in Step 1, we examine how these models are declared within each repository. Specifically, we perform textual searches using the identified model names across relevant project artifacts, including configuration files (\eg \texttt{YAML}, \texttt{JSON}) and the README. This step allows us to assess whether PTM dependencies are explicitly and centrally documented, or whether such declarations are scattered or implicit. 

\paragraph{Step 3: Identify PTM Version Declarations (for RQ1.2).} 

\noindent Versioning promotes the reproducibility, stability, and traceability of ML experiments. Since pretrained models may change over time (\eg due to updated weights, bug fixes, or architectural changes), specifying the version of a PTM used in a project helps guarantee consistent behavior across environments and over time.

Hugging Face models are hosted in versioned Git repositories. Users can specify a particular model version using the \texttt{revision} parameter in the \texttt{from\_\allowbreak pretrained()} API, for example:

\begin{center}
\texttt{AutoModel.from\_pretrained("bert-base-uncased", revision="v4.2.0")}
\end{center}

Similarly, models loaded from PyTorch Hub can reference specific commits or tags by specifying the appropriate version string:

\begin{center}
\texttt{torch.hub.load("pytorch/vision:v0.10.0", "resnet18")}
\end{center}

To identify version declarations, we statically analyze model-loading code and check whether these version parameters (\eg \texttt{revision} in Hugging Face, version tags in PyTorch Hub) are explicitly provided. 
\new{In addition to source code, we also analyze configuration files (\eg YAML and JSON files) and project README files to identify explicit references to PTM versions.} This allows us to measure how often projects declare exact PTM versions, and thus, how they approach reproducibility in their dependency practices.

\subsubsection{Results}
\label{subsec:rq1_results}

\new{
\noindent \textbf{Prevalence of Model Families.}}
\new{
The sampled 401 projects exhibit several well-represented model families rather than a single dominant group. Across the 381 unique PTMs
identified, we find 79 base model families. The most represented family is \texttt{BERT} (12.1\%, 46/381), including models such as
\texttt{bert-base-uncased}, \texttt{roberta-base}, and \texttt{distilbert-base-uncased}. \texttt{ResNet} follows with 10.5\%(40/381), including \texttt{resnet50}, \texttt{resnet101}, and \texttt{resnext50\_32x4d}. \texttt{ViT} accounts for 9.2\% (35/381), with examples such as \texttt{vit\_base\_patch16\_224}, \texttt{ViT-B-16}, and \texttt{dino\_vitb16}. \texttt{Stable Diffusion} contributes 7.1\% (27/381), including \texttt{stable-diffusion-v1-5} and \texttt{stable-diffusion-xl-base-1.0}, while the \texttt{GPT} family represents 5.5\% (21/381), including \texttt{gpt2}, \texttt{gpt2-large}, and \texttt{EleutherAI/gpt-neo-2.7B}. Other notable families include \texttt{CLIP} (4.5\%), \texttt{T5} (4.2\%), \texttt{Wav2Vec} 2.0(3.4\%), \texttt{VGG} (2.4\%), \texttt{BART} (2.1\%), and \texttt{MobileNet} (1.6\%). Overall, the distribution reflects a broad and diverse ecosystem spanning vision, language, speech, and generative models.
}

\begin{figure}[htb]
    \centering
    \includegraphics[width=0.9\linewidth]{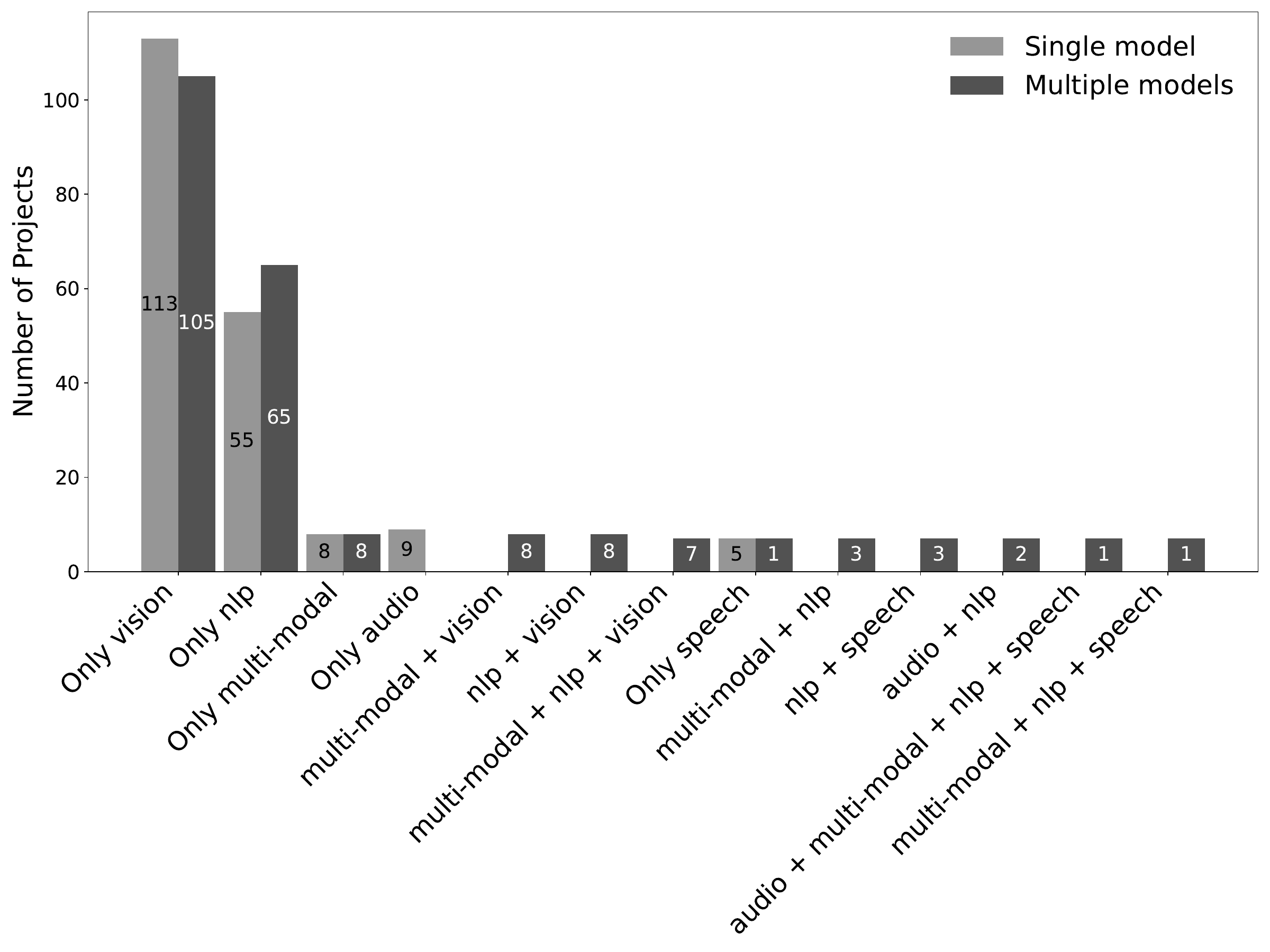}
    \caption{Number of projects by modality combinations.}
    \label{fig:repo_types}
\end{figure}

\vspace{0.1cm}
\noindent \textbf{PTM Dependency Structure.}
\new{To characterize how OSS projects organize their PTM dependencies, we consider the number of PTMs integrated within a project (single vs.\ multiple) and the modality and architectural type of the models (base model family) involved. These dimensions jointly capture structural complexity at different levels.}
Figure~\ref{fig:repo_types} summarizes the distribution of target projects, grouped by their model type. Most single-PTM projects rely exclusively on vision models (113 projects) or NLP (55 projects), with only a handful centered on speech, audio, or multi-modal models. By contrast, multi-PTM projects overwhelmingly involve cross-modality combinations, most commonly NLP + vision (8 projects). Other combinations, such as multi-modal + vision (8 projects), multi-modal + NLP (3 projects), or NLP + speech (3 projects), are less frequent but illustrate the diversity of integration patterns. Multiple-PTM projects tend to combine models across different modalities.

Within the Multiple PTMs category, we further distinguish two structural subtypes: \textit{Interchangeable} and \textit{Complementary}. 
Figure~\ref{fig:ptm_dependency_tree} shows the distribution of these dependency structures in the examined project set.

\begin{figure}[h]
\centering
\begin{adjustbox}{cfbox=black 1pt 3pt}

\begin{tikzpicture}[
  sibling distance=10em,
  level distance=4em,
  every node/.style = {shape=rectangle, draw, align=center, font=\small, rounded corners=2pt}
  ]
\node {PTM Usage\\401 projects}
  child { node {Single PTM\\193 (48\%)}
  }
  child { node {Multiple PTMs\\208 (52\%)}
    child { node {Interchangeable\\149\\(37\% of all 401)} }
    child { node {Complementary\\92\\(23\% of all 401)} }
  };
\end{tikzpicture}
\end{adjustbox}

\caption{\new{PTM dependency structures in OSS projects. The top-level
percentages (48\%, 52\%) represent proportions across all 401 projects.
For each subcategory, we report both the count and its percentage
relative to the full set of 401 projects (rather than only multi-PTM
projects). Some multi-PTM projects contain both interchangeable and
complementary PTMs; therefore, the subcategory totals are not mutually
exclusive (149 + 92 $>$ 208).}}
\label{fig:ptm_dependency_tree}
\end{figure}

\textit{Interchangeable PTMs structure} exists when multiple PTMs are included in a project to fulfill the same functional role, forming parallel and substitutable dependency structure. For instance, a project might support variants such as \texttt{clip\_vit}, \texttt{clip\_vit\_b16}, and \texttt{clip\_rn50}, all used for the same task (\eg vision encoding). These models typically belong to the same architecture family and differ in size, configuration, or pretraining dataset. Their usage reflects a flexible dependency structure, where models can be swapped at runtime based on performance or resource constraints, without requiring changes to the surrounding pipeline.

\textit{Complementary PTMs structure} exists when a project integrates multiple PTMs with multiple roles, forming a functionally diverse dependency structure. Each PTM is typically used for a different task or type of data (modality). For example, a project might use \texttt{Wav2Vec2} for speech input, \texttt{BERT} for textual encoding. These PTMs are non-substitutable and together enable multi-modal or multi-stage processing. Complementarity may also occur within a single modality if models address different subtasks. This structure reflects intentional modular design, where each PTM dependency supports a specialized component of the system.

In our dataset, interchangeable PTMs are typically drawn from the same family, while complementary PTMs typically combine models from different families to address different tasks or modalities. This observation is supported by a manual review of \new{135 randomly sampled multi-PTM projects out of 208 multi-PTM projects, where the sample size was determined to achieve a 95\% confidence level with a 5\% margin of error.}
Our results show that \new{76\%} of the projects used PTMs from the same family in interchangeable roles and \new{86\%} of the projects combined PTMs from different families in complementary roles. \new{Despite these dominant patterns, the mapping between family membership and functional role is not strict. Some projects use models from different families interchangeably for the same task, while others employ multiple models from the same family in complementary roles. For instance, \texttt{UMR}\footnote{\url{https://github.com/NVlabs/UMR}} implements a configurable perceptual network that accepts \texttt{vgg16}, \texttt{alexnet}, \texttt{resnet}, or \texttt{squeezenet} as a runtime parameter: all serving the identical role of computing perceptual similarity, despite belonging to different families. Conversely, \texttt{lama-cleaner}\footnote{\url{https://github.com/sanster/lama-cleaner}} uses multiple Stable Diffusion variants not as interchangeable
alternatives, but in distinct stages of its image inpainting pipeline.}

These two dependency structures reflect different design goals: interchangeable PTMs typically offer runtime flexibility and configurability, often within the same family but occasionally across families; complementary PTMs typically enable functional decomposition through modular, multi-modal, or multi-stage architectures, often across families but occasionally within one. A comparison of their typical characteristics is presented in Table~\ref{tab:ptm-usage-comparison}.

\begin{table}[htb]
\centering
\caption{Comparison of interchangeable and complementary PTM dependency structures}
\label{tab:ptm-usage-comparison}
\renewcommand{\arraystretch}{1.3} 
\begin{tabular}{>{\raggedright\arraybackslash}p{2.6cm} >{\raggedright\arraybackslash}p{4cm} >{\raggedright\arraybackslash}p{4cm}}
\toprule
\textbf{Aspect} & \textbf{Interchangeable PTMs} & \textbf{Complementary PTMs} \\
\midrule
Architectural similarity & 
Typically from the same base model family(\eg CLIP variants, ViT models) & 
Typically from different families (\eg BERT, Wav2Vec2, ResNet); may also differ within a modality \\

Configuration differences & 
Vary in size, tokenizer, or pretraining dataset & 
Vary in both configuration and architecture, tailored to specific tasks \\

Modality & 
Same modality (\eg all text, all vision, or all multi-modal) & 
Same or cross-modality (\eg combining text, vision, speech) \\

Functional role & 
Serve the same task; models are used interchangeably & 
Generally serve distinct functions; models are used jointly to form a complete workflow \\
\bottomrule
\end{tabular}
\end{table}

\vspace{0.5em}

\noindent \textbf{PTM Dependency Documentation.}  
Our analysis revealed three high-level PTM dependency patterns. The first pattern is defined by projects with no PTM declarations outside of the source code. The second pattern includes projects with partial or scattered mentions. The third pattern covers projects with explicit and centralized documentation.

\begin{figure}[ht]
    \centering
    \includegraphics[width=0.98\linewidth]{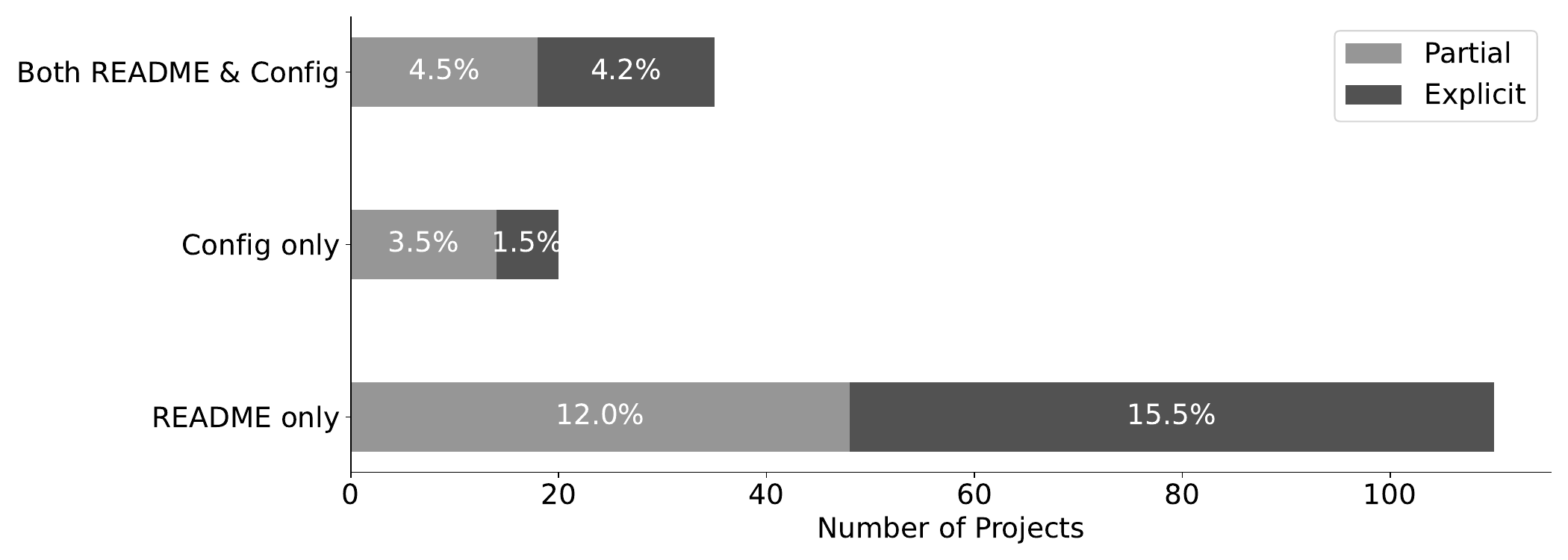}
    \captionsetup{skip=3pt} 
        \caption{
        Breakdown of PTM mentions in projects. Partial mentions refer to projects that mention some PTMs either in documentation or configuration files but not comprehensively. Explicit mentions refer to projects that fully declare all used PTMs in external files. Subcategories show projects documenting PTMs only, configuring PTMs only, or including both, with counts and percentages for each.
        }

    \label{fig:breakdown}
\end{figure}

Among the projects, 58.9\% (236 projects) mention PTMs only in source code and not in configuration or documentation files. Identifying the actual PTM dependencies in these cases requires code analysis, especially for models loaded dynamically at runtime. Another 20.0\% (80 projects) provide partial mentions of PTMs. Figure~\ref{fig:breakdown} details this breakdown, showing that 12.0\% (48 projects) mention PTMs only in documentation, 3.5\% (14 projects) only in configuration files, and 4.5\% (18 projects) in both. These mentions are often scattered or informal, such as listing model names without version information or source specification.
Only \ProjectsCentralizedDocumentation (85 projects) explicitly declare all used PTMs in external files. Of these, 15.5\% (62 projects) do so only in documentation, 1.5\% (6 projects) only in configuration files, and 4.2\% (17 projects) in both.

Our review also found that even when PTM names are present in configuration files (\eg YAML or JSON), they rarely include explicit hub information, such as Hugging Face or PyTorch Hub, which limits reproducibility and automation.

\vspace{0.1cm}
\noindent \new{\textbf{Documentation Practices in Single-PTM vs. Multi-PTM Projects.} To understand how project complexity affects documentation practices, we analyzed whether single-PTM and multi-PTM projects differ in their dependency documentation approaches. Among single-PTM projects, we found that 70.5\% (136/193) of single-PTM projects mention their PTM only in source code, while 29.5\% (57/193) explicitly declare it in outside source code files: 20.7\% (40 projects) in README only, 2.6\% (5 projects) in configuration files only, and 6.2\% (12 projects) in both.}

\new{In contrast, multi-PTM projects introduce a third category: partial documentation, where only some PTM dependencies are documented outside the source code. Among multi-PTM projects, 48.1\% (100/208) mention PTMs only in source code, and 38.5\% (80/208) provide partial mentions: 23.1\% (48 projects) in README only, 6.7\% (14 projects) in configuration files only, and 8.7\% (18 projects) in both. Only 13.5\% (28/208) explicitly declare \textit{all} their PTM dependencies in external files: 8.2\% (17 projects) in README only, 0.5\% (1 project) in configuration files only, and 4.8\% (10 projects) in both. While multi-PTM projects more frequently \textit{begin} documenting their dependencies outside source code (51.9\% vs.\ 29.5\%), they rarely achieve \textit{complete} coverage (13.5\% vs.\ 29.5\%), suggesting that increasing project complexity motivates documentation efforts but simultaneously makes comprehensive documentation harder to maintain.
}

\vspace{0.1cm}
\noindent \textbf{Versioning Practices.}
Version control is critical for ensuring reproducibility, especially as PTMs
may evolve over time with updated weights or architecture refinements~\citep{ajibode2025towards}. 
However, we find that explicit PTM versioning is rare. Out of 401 projects,
only 47 (12\%) specify a model version, either via the revision parameter in
Hugging Face’s \texttt{from\_pretrained()} API or a version tag in
\texttt{torch.hub.load()}. These declarations appear exclusively in code; we
did not find any PTM version information in configuration or documentation
files. This lack of structured versioning further limits traceability and long-term reliability in PTM-based development. These findings highlight a gap in standardized dependency documentation practices for PTMs and suggest the need for better tools and conventions to promote reproducibility and transparency in ML development.

\FindingsBox{RQ1: \rqone}{
Our analysis reveals varied and non-uniform patterns in how projects use and declare dependencies on PTMs. Among the 401 projects studied, \ProjectsMultiPTM utilize more than one PTM. Within these multi-PTM projects, some dependencies are interchangeable (37\%), allowing one PTM to substitute for another in the workflow, while others are complementary (23\%), where PTMs serve distinct functional roles. We also observe inconsistency in dependency declaration practices: PTM references are frequently decentralized, \ie appearing across source code, documentation, and configuration files (\eg config.yaml, config.json). Only \ProjectsCentralizedDocumentation of projects declares all PTMs in a centralized file outside of code. Versioning practices are similarly limited: only 12\% specify a model version, typically in code.
}

\subsection{\textbf{RQ2: \rqtwo}}
\label{subsec:rq2}

To answer RQ2, we analyze both the \textit{stages} involved in PTM reuse pipelines and the \textit{organization} of these stages. Our analysis is based on a manual coding of the \TotalProjects projects described in \S\ref{subsec:data_collection}. Specifically, we first identify the sequence of ML stages that involve PTMs, and then examine how these stages are connected to form complete PTM reuse pipelines.

\vspace{0.1cm}

\subsubsection{Methodology}
\label{subsec:rq2_methodology}

\paragraph{Step 1: Stage Identification.}
Conventional ML pipelines typically follow a data-first sequence, progressing from data acquisition to model building and training (\S\ref{fig:background_conv_pipeline}). In contrast, PTM reuse pipelines adopt a reuse-first organization: developers often begin by loading external PTMs and adapting them to specific tasks, instead of building and training models from scratch. This shift moves modeling and training away from being central activities and instead emphasizes new tasks such as prompt engineering, adapter integration, fine-tuning, and post-inference handling.

Despite these fundamental differences, the literature lacks a clear definition of PTM reuse as a cohesive pipeline. As outlined in Section~\ref{sec:background}, we define a PTM reuse pipeline as a structured workflow describing how PTMs are retrieved, adapted, and operationalized. However, the specific stages within such pipelines remain unclear. To address this gap, we conduct a targeted literature review with the specific goal of identifying PTM reuse pipeline stages. Following the principles of systematic mapping studies in software engineering~\citep{biswas2022art}, we query academic sources using the search string: 

\begin{quote}
\hspace*{2em}\texttt{("machine learning pipeline" OR "data science workflow" OR "lifecycle")} \\
\hspace*{2em}\texttt{AND ("pre-trained model" OR "foundation model" OR "large language model")} \\
\hspace*{2em}\texttt{AND (published in 2018--2024)}
\end{quote}

\new{We derived the search keywords from established ML workflow terminology
(\eg pipeline, workflow, lifecycle) together with PTM-specific concepts
(\eg pre-trained model, foundation model, large language model). During the
scoping phase, we evaluated additional candidate terms, such as ``transfer
learning pipeline'' and ``fine-tuning pipeline.'' However, these queries
primarily retrieved studies unrelated to ML pipeline design and yielded no
relevant work on PTM reuse pipelines; therefore, they were excluded from the
final keyword set. We restricted the search to the 2018–2024 period to capture the emergence and rapid adoption of large-scale PTMs following the introduction of BERT.}

The search was conducted across IEEE Xplore, ACM Digital Library, and
Google Scholar (queried using the Publish or Perish tool\footnote{See
\url{https://harzing.com/resources/publish-or-perish}}). \new{The
initial search returned 900 papers. After removing duplicates and
non-archival proceedings entries (e.g., editorials, workshop summaries,
and short front-matter without full methodological descriptions), 491
papers remained. We excluded such entries because they typically lack
sufficient technical detail to analyze PTM reuse pipelines and often
overlap with full peer-reviewed publications. The first author then
conducted title/abstract screening. Papers were excluded if they focused
solely on algorithms without pipeline context, discussed PTMs
conceptually without a workflow perspective, or addressed
DevOps/governance rather than model-level reuse. Ambiguous cases were
discussed with co-authors until consensus.} After removing duplicates and irrelevant papers based on titles and abstracts, 300 papers are selected for full-text review.

Our review finds no prior work provides a taxonomy of PTM reuse pipelines or their stages. Most papers describe isolated PTM reuse activities (\eg fine-tuning, prompt engineering) without situating them in a unified pipeline structure. \new{We developed our taxonomy through open coding. We started with a preliminary API dictionary from prior ML pipeline work and PTM frameworks (Hugging Face, PyTorch), identifying characteristic functions like \texttt{from\_pretrained()}, \texttt{train()}, \texttt{evaluate()}. We used contextual cues (mode switches, optimizer presence, output usage) to resolve ambiguities where the same API appears in multiple stages. The first author developed the initial taxonomy, and all authors participated in three rounds of discussion to refine stage definitions.} Based on our analysis, we manually identify a minimal, reusable set of PTM-centric stages that consistently appear across both the literature and open-source projects. We focus specifically on model-level, code-observable practices, and intentionally exclude full-stack, DevOps, or governance frameworks~\citep{fitsilis2024dollmc}. Table~\ref{tab:merged_pipeline_stages} lists the eight identified stages. Stages such as data processing and post-processing are treated at a coarse level, as their fine-grained substeps (\eg data acquisition, data preparation) vary widely and are often implicit or domain-specific.

\vspace{1em}

To map these stages to real-world code, we follow an approach inspired by prior ML pipeline analyses~\citep{wang2021restoring,biswas2022art}, where framework-specific API calls are mapped to conceptual stages. However, because existing taxonomies are not tailored to PTM reuse, we extend this process with open coding to inductively identify new reuse-specific stages based on emerging patterns.

We begin with a preliminary API dictionary constructed from prior literature and enriched with PTM-related frameworks (\eg Hugging Face, PyTorch). This includes characteristic functions such as \texttt{from\_pretrained()}, \texttt{Trainer.train()}, \texttt{evaluate()}, and \texttt{predict()}.

Because APIs like \texttt{model()} or \texttt{forward()} are reused across stages, we resolve ambiguities using contextual cues such as:
\begin{itemize}
\item Mode switches (\eg \texttt{model.train()} vs. \texttt{model.eval()}),
\item Optimization behavior (\eg presence of \texttt{optimizer.step()}),
\item Functional use of outputs (\eg loss computation vs. prediction).
\end{itemize}

\new{The labeling process consisted of two steps. In the first step, the first author labeled all projects, and all authors participated in three rounds of discussion to resolve ambiguous cases and refine the stage definitions. In the second step, we conducted a cross-check by having one co-author independently annotate a randomly selected subset of 282 PTMs based on the finalized stage definitions, without access to the first author's labels. This subset is generated using a random sampling approach calculated to achieve a 95\% confidence level and a 5\% margin of error, ensuring it is statistically representative of the full dataset (\ie 1,059 pipeline instances), resulting in an inter-rater agreement score of 0.81, indicating substantial agreement~\citep{viera2005understanding}.} 

\paragraph{Step 2: Identifying PTM Adaptation Strategies in Practice (for RQ2.1).}
Because adaptation is a defining feature of PTM reuse, we conduct an in-depth analysis of the strategies developers employ during the Model Adaptation stage of the PTM reuse pipeline. We first extracted the default configurations of the PTMs used in our dataset projects from their official documentation, including (1) Model Architecture: the structural layout, such as layer types and depths, and (2) Parameter Settings: the number of parameters and their default trainable or frozen status. These defaults served as a baseline for the initial state of each PTM before any user intervention. We then compared these defaults to actual usage observed in projects to detect modifications.

\paragraph{Step 3: Categorizing organizations of Stages (for RQ2.2).}
After identifying the individual stages involved in PTM reuse pipeline (Step 1), we focus on how these stages are structured and interconnected, which we refer to as the \textit{organization} of stages within a pipeline.
To characterize these organizations, we analyzed patterns across annotated projects from Step 1, examining both the sequence of stages and the functional role of the PTM at each stage. We iteratively grouped similar workflows and refined our categorization scheme through collaborative discussion, using representative examples from our dataset to guide decisions.

\subsubsection{Results}
\label{subsec:rq2_result}


\vspace{0.5em}
\noindent \textbf{Stages in PTM Reuse Pipelines.} 
From our empirical study of \TotalProjects target projects, we derive a representative taxonomy of PTM reuse pipelines consisting of 10 distinct stages, as summarized in Table~\ref{tab:merged_pipeline_stages}. These stages span the entire lifecycle of PTM integration, from model initialization to downstream delivery, providing a conceptual structure for analyzing PTM reuse practices.


{
\small
\renewcommand{\arraystretch}{1.1}
\begin{landscape}
\begin{table*}[ht]
\centering
\caption{Unified taxonomy of PTM reuse pipeline stages from our empirical study of \TotalProjects projects, arranged chronologically from model initialization to final delivery. Stages that differ from or expand on prior literature are highlighted in \textbf{bold}, including \textbf{Adaptation (ADPT)}, \textbf{\new{Prompting (PROM)}}, \textbf{Post-Processing (POST)}, and the newly surfaced \textbf{Evaluation (EVAL)} and \textbf{Delivery (DLV)} stages. Literature-based highlights and our taxonomy-based descriptions are provided for each stage.
}
\label{tab:merged_pipeline_stages}
\begin{tabular}{p{0.15\linewidth} p{0.35\linewidth} p{0.35\linewidth}}
      \toprule
      \textbf{Stage} & \textbf{Literature Highlights and Notes} & \textbf{Description (Taxonomy)} \\
      \midrule
      Model Initialization (INIT) & Loading pretrained weights and architectures from repositories or local storage~\citep{ramachandran2023will, ran2025foundation}. & Loads a pretrained model from local storage or external repositories. Restores architecture and parameters for inference or adaptation. \\[4pt]

      \textbf{Model Adaptation (ADPT)} & Integrating task-specific heads/modules including LoRA~\citep{ramachandran2023will, davies2024social}. & Adjusts or customizes architecture or parameters without extra training. \\[4pt]

      Data Processing (PROC) & Preparing inputs (\eg normalization). Often coarse-grained in literature~\citep{amershi2019software, ran2025foundation}. & Converts raw data into formats compatible with pretrained models, including tokenization, resizing, cleaning, augmentation. \\[4pt]

      \textbf{\new{Prompting (PROM)}} & Crafting prompts—manually or automatically—to guide generative PTMs~\citep{ramachandran2023will}. & Creates prompts manually, templated, or automatically using prompt engineering techniques. \\[4pt]

      Feature Engineering (FEAT) & Constructing features from raw data. Often bypassed due to pretrained embeddings~\citep{biswas2022art}. & Transforms raw data into structured, informative features for retrieval, classification, or other tasks. \\[4pt]

      Fine-Tuning (FT) & Further training PTMs on labeled or domain-specific data~\citep{ramachandran2023will, ran2025foundation, davies2024social}. & Trains the model on additional labeled data to specialize it for downstream tasks. \\[4pt]

      Inference (INF) & Running the model on new inputs (\eg sequence generation, classification)~\citep{ramachandran2023will, xia2024towards}. & Produces sequences for generative models or class labels/scores for discriminative tasks. \\[4pt]

      \textbf{Post-Processing (POST)} & Formatting or filtering outputs (\eg detokenization, label mapping)~\citep{biswas2022art}. & Refines outputs for usability, including filtering, hallucination detection, re-ranking. \\[4pt]
      \midrule
      \textbf{Evaluation (EVAL)} & -- & Assesses outputs using quantitative metrics or qualitative criteria. \\[4pt]

      \textbf{Delivery (DLV)} & -- & Provides outputs via APIs, dashboards, or other integration points. \\[4pt]
      \bottomrule
    \end{tabular}%
\end{table*}
\end{landscape}

}

Table \ref{tab:merged_pipeline_stages} shows that while many stages in our taxonomy mirror those in the literature-based taxonomy, our analysis of real-world code surfaced two additional stages: 
\begin{enumerate}
\setlength{\itemindent}{2em} 

\item \textbf{Evaluation (EVAL)} assesses model outputs systematically using quantitative metrics 
(\eg FID score, BLEU, accuracy) or qualitative criteria to measure quality and task alignment. \new{Note that we introduce \textit{Evaluation (EVAL)} as a distinct stage in the PTM reuse pipeline (Table.~\ref{tab:merged_pipeline_stages}), even though conventional ML pipelines (Fig.~\ref{fig:background_conv_pipeline}) also include an evaluation step, because the two evaluation stages differ in purpose and operational context. In conventional ML pipelines, evaluation typically refers to validation or testing during model development, where models trained from scratch are assessed to guide improvement or selection prior to deployment. In contrast, the \textit{Evaluation (EVAL)} in PTM reuse pipeline denotes post-inference assessment of a ready-to-use model (an existing PTM or adapted/fine-tuned PTM)’s outputs (e.g., quality, safety, or consistency) during application use rather than during model development.}

\item \textbf{Delivery (DLV)} provides final outputs via APIs, interfaces, dashboards, 
or other integration points for practical deployment. It exposes the results of the PTM reuse pipeline via interfaces such as APIs, web dashboards, or other user- or system-facing endpoints. This stage focuses on making the model’s outputs accessible and usable, providing a bridge between the pipeline’s computations and practical consumption.
\end{enumerate}

\begin{table}[h]
\centering
\caption{
    Adaptation strategies for using PTMs. 
    \textcolor{blue}{Addition} and \textcolor{red}{Deletion} operations are highlighted accordingly.
}
\label{tab:ptm_adaptation}
\renewcommand{\arraystretch}{1.3}
\resizebox{\textwidth}{!}{%
\begin{tabular}{p{3.8cm} p{7.5cm}}
\toprule
\textbf{Adaptation} & \textbf{Description and Example} \\
\midrule

Using PTM as-is & The PTM is applied directly to a downstream task without modifications. 
\newline \textit{Example: Using ResNet-50 pretrained on ImageNet for a similar image classification task.} \\

Using PTM weights & \new{Pretrained weights are used to \textbf{initialize} a custom model structure. This involves mapping parameters onto a new architecture.
\newline \textit{Example: Transferring ResNet-50 weights to a custom backbone for human mesh recovery.~\href{https://github.com/YuliangXiu/ICON} }} \\
\textcolor{blue}{Addition} & \new{New structural components are appended to the PTM backbone to extend its functionality.}
\newline \textit{Example: Adding a classification head on top of BERT for intent recognition.} \\

\textcolor{red}{Deletion} & Parts of the PTM are removed to simplify or repurpose the model.
\newline \textbf{\textcolor{red}{Head Deletion Only}:} Final classifier is removed; the PTM acts as a feature extractor.
\newline \textit{Example: Using all convolutional layers of ResNet for image retrieval.}
\newline \textbf{\textcolor{red}{Head + Layer Deletion}:} Both the classifier head and some upper layers are removed to reduce depth.
\newline \textit{Example: Extracting features from early layers of AlexNet for texture matching.} \\

Modification & The final layer is replaced with a task-specific component, LoRA, Prefix tuning, Adapter layers.
\newline \textit{Example: Substituting BERT’s pooled output head with a regression layer.} \\

Configuration changes / optimization & Modifications to dropout, hidden size, or other hyperparameters to improve performance or fit.
\newline \textit{Example: Adjusting dropout in a BERT variant to better suit low-resource settings.} \\
\bottomrule
\end{tabular}
}
\label{tab:ptm_adaptation}
\end{table}


In addition, several existing stages take on different emphases in practice. 
For example, \textbf{Adaptation} appears prominently in real-world pipelines, 
encompassing not only the addition of a head but also head replacement, 
selective layer use, architectural changes, and configuration tweaks (Table~\ref{tab:ptm_adaptation}). 
Fine-tuning itself spans a spectrum of strategies, 
from fully updating all parameters to partially freezing layers
(\eg freezing lower-level feature extractors while training the higher-level layers), thereby balancing computational efficiency with task specialization. 
Likewise, \textbf{Prompting} and \textbf{Post-Processing} often serve as critical operational stages in generative tasks, with the latter ensuring output quality and safety.  

Compared to prior work on traditional ML, foundation models, and large language models~\citep[\eg][]{ramachandran2023will, ran2025foundation}, our taxonomy is model-type-agnostic and captures a broader range of strategies observed in empirical PTM reuse.

\vspace{0.5em}

\begin{figure}
\begin{adjustbox}{cfbox=black 1pt 3pt} 
    \begingroup
    \LARGE
    \resizebox{0.95\textwidth}{!}{%
        \begin{tikzpicture}[
    box/.style={rectangle, draw, rounded corners=4pt, minimum height=0.7cm, inner sep=4pt, font=\small},
    optbox/.style={rectangle, draw, dashed, rounded corners=4pt, minimum height=0.7cm, inner sep=4pt, font=\small},
    orbox/.style={rectangle, draw, rounded corners=4pt, minimum height=0.7cm, inner sep=4pt, font=\small, fill=gray!20},
    edge/.style={thick},
    mand/.style={circle, fill=black, inner sep=2pt},
    opt/.style={circle, draw, thick, fill=white, inner sep=2pt},
]

\def\xA{0}    
\def\xB{6}    
\def\xC{12}   

Shared y positions for all 10 rows
\def\yMI{-0.9}    
\def\yMA{-1.7}    
\def\yDP{-2.5}    
\def\yPR{-3.3}    
\def\yFE{-4.9}    
\def\yFT{-5.7}    
\def\yIN{-6.5}    
\def\yPP{-7.3}    
\def\yEV{-8.1}    
\def\yDL{-8.9}    


\node[box, font=\small\bfseries] (titleA) at (\xA, 0) {Feature Extraction};
\node[box, font=\small\bfseries] (titleB) at (\xB, 0) {Generative};
\node[box, font=\small\bfseries] (titleC) at (\xC, 0) {Discriminative};


\draw[edge] (titleA.south) -- ++(0,-0.3) -- ++(0,-8.5);
\draw[edge] (titleB.south) -- ++(0,-0.3) -- ++(0,-8.5);
\draw[edge] (titleC.south) -- ++(0,-0.3) -- ++(0,-8.5);


\draw[edge] (\xA,\yMI) -- ++(0.5,0); \node[mand] at (\xA+0.1,\yMI) {};
\node[box, anchor=west] at (\xA+0.5,\yMI) {Model Initialization};

\draw[edge] (\xB,\yMI) -- ++(0.5,0); \node[mand] at (\xB+0.1,\yMI) {};
\node[box, anchor=west] at (\xB+0.5,\yMI) {Model Initialization};

\draw[edge] (\xC,\yMI) -- ++(0.5,0); \node[mand] at (\xC+0.1,\yMI) {};
\node[box, anchor=west] at (\xC+0.5,\yMI) {Model Initialization};


\draw[edge] (\xA,\yMA) -- ++(0.5,0); \node[mand] at (\xA+0.1,\yMA) {};
\node[box, anchor=west] at (\xA+0.5,\yMA) {Model Adaptation};

\draw[edge] (\xB,\yMA) -- ++(0.5,0); \node[mand] at (\xB+0.1,\yMA) {};
\node[box, anchor=west] at (\xB+0.5,\yMA) {Model Adaptation};

\draw[edge] (\xC,\yMA) -- ++(0.5,0); \node[mand] at (\xC+0.1,\yMA) {};
\node[box, anchor=west] at (\xC+0.5,\yMA) {Model Adaptation};


\draw[edge] (\xA,\yDP) -- ++(0.5,0); \node[mand] at (\xA+0.1,\yDP) {};
\node[box, anchor=west] at (\xA+0.5,\yDP) {Data Processing};

\draw[edge] (\xB,\yDP) -- ++(0.5,0); \node[mand] at (\xB+0.1,\yDP) {};
\node[box, anchor=west] at (\xB+0.5,\yDP) {Data Processing};

\draw[edge] (\xC,\yDP) -- ++(0.5,0); \node[mand] at (\xC+0.1,\yDP) {};
\node[box, anchor=west] at (\xC+0.5,\yDP) {Data Processing};



\draw[edge] (\xB,\yPR) -- ++(0.5,0); \node[mand] at (\xB+0.1,\yPR) {};
\node[box, anchor=west] at (\xB+0.5,\yPR) {Prompt};


\draw[edge] (\xA,\yFE) -- ++(0.5,0); \node[opt] at (\xA+0.1,\yFE) {};
\node[optbox, anchor=west] at (\xA+0.5,\yFE) {Feature Engineering};

\draw[edge] (\xB,\yFE) -- ++(0.5,0); \node[opt] at (\xB+0.1,\yFE) {};
\node[optbox, anchor=west] at (\xB+0.5,\yFE) {Feature Engineering};

\draw[edge] (\xC,\yFE) -- ++(0.5,0); \node[opt] at (\xC+0.1,\yFE) {};
\node[optbox, anchor=west] at (\xC+0.5,\yFE) {Feature Engineering};


\draw[edge] (\xA,\yFT) -- ++(0.5,0);
\node[orbox, anchor=west] at (\xA+0.5,\yFT) {Fine-Tuning};

\draw[edge] (\xB,\yFT) -- ++(0.5,0);
\node[orbox, anchor=west] at (\xB+0.5,\yFT) {Fine-Tuning};

\draw[edge] (\xC,\yFT) -- ++(0.5,0);
\node[orbox, anchor=west] at (\xC+0.5,\yFT) {Fine-Tuning};


\draw[edge] (\xA,\yIN) -- ++(0.5,0);
\node[orbox, anchor=west] at (\xA+0.5,\yIN) {Inference};

\draw[edge] (\xB,\yIN) -- ++(0.5,0);
\node[orbox, anchor=west] at (\xB+0.5,\yIN) {Inference};

\draw[edge] (\xC,\yIN) -- ++(0.5,0);
\node[orbox, anchor=west] at (\xC+0.5,\yIN) {Inference};

\draw[thick, decorate, decoration={brace, amplitude=4pt, mirror}]
    (\xA+0.3,-5.4) -- (\xA+0.3,-6.7) node[midway, left=6pt, font=\scriptsize\bfseries] {OR};
\draw[thick, decorate, decoration={brace, amplitude=4pt, mirror}]
    (\xB+0.3,-5.4) -- (\xB+0.3,-6.7) node[midway, left=6pt, font=\scriptsize\bfseries] {OR};
\draw[thick, decorate, decoration={brace, amplitude=4pt, mirror}]
    (\xC+0.3,-5.4) -- (\xC+0.3,-6.7) node[midway, left=6pt, font=\scriptsize\bfseries] {OR};


\draw[edge] (\xA,\yPP) -- ++(0.5,0); \node[opt] at (\xA+0.1,\yPP) {};
\node[optbox, anchor=west] at (\xA+0.5,\yPP) {Post-Processing};

\draw[edge] (\xB,\yPP) -- ++(0.5,0); \node[opt] at (\xB+0.1,\yPP) {};
\node[optbox, anchor=west] at (\xB+0.5,\yPP) {Post-Processing};

\draw[edge] (\xC,\yPP) -- ++(0.5,0); \node[opt] at (\xC+0.1,\yPP) {};
\node[optbox, anchor=west] at (\xC+0.5,\yPP) {Post-Processing};



\draw[edge] (\xB,\yEV) -- ++(0.5,0); \node[opt] at (\xB+0.1,\yEV) {};
\node[optbox, anchor=west] at (\xB+0.5,\yEV) {Evaluation};

\draw[edge] (\xC,\yEV) -- ++(0.5,0); \node[opt] at (\xC+0.1,\yEV) {};
\node[optbox, anchor=west] at (\xC+0.5,\yEV) {Evaluation};



\draw[edge] (\xB,\yDL) -- ++(0.5,0); \node[opt] at (\xB+0.1,\yDL) {};
\node[optbox, anchor=west] at (\xB+0.5,\yDL) {Delivery};

\draw[edge] (\xC,\yDL) -- ++(0.5,0); \node[opt] at (\xC+0.1,\yDL) {};
\node[optbox, anchor=west] at (\xC+0.5,\yDL) {Delivery};

\node[anchor=north west, font=\small\bfseries, text=black!70] at (0,-9.2) {Legend:};
\node[mand, label={[font=\scriptsize]right:Mandatory}] at (0.2,-10.0) {};
\node[opt, label={[font=\scriptsize]right:Optional}] at (0.2,-10.5) {};
\node[optbox, minimum height=0.3cm, minimum width=0.5cm] at (1.6,-10.5) {};
\node[orbox, minimum height=0.3cm, minimum width=0.5cm, label={[font=\scriptsize]right:Or ($\geq$1)}] at (0.2,-11.0) {};

\end{tikzpicture}
    }
    \endgroup
\end{adjustbox}
\caption{\new{Feature model of the PTM Reuse Pipeline variability. Mandatory stages (solid) are shared across all pipeline types. Dashed boxes indicate optional stages. Grey boxes denote an or-group, i.e., at least one of Fine-Tuning or Inference must be present. Cross-tree constraints capture stage availability per pipeline type.}
}
\label{fig:pipeline_orientations}
\end{figure}

\noindent \textbf{Organization of the Stages in PTM Reuse Pipelines.}
To characterize the structure of pretrained model (PTM) reuse workflows, we identify three dominant pipeline organizations based on the functional role of the PTM. These organizations, illustrated in Figure~\ref{fig:pipeline_orientations}, capture whether the PTM is primarily used as a feature extractor, a generative engine, or a discriminative predictor. While many stages (\eg model initialization, adaptation) appear in pipelines, their function and sequencing differ depending on the intended use of the PTM.

Each pipeline includes at least one of the two stages, \textit{inference} or \textit{fine-tuning}, which may appear independently or together, depending on the level of task alignment required. Inference is the most consistently used stage, appearing in around 75\% of pipelines. Some stages (\eg post-processing, feature engineering) are used selectively, depending on downstream requirements.

\vspace{0.5em}
\noindent The three pipeline organizations are:

\begin{enumerate}
\item \textbf{Feature Extraction–Oriented Pipeline.}
In this organization, the PTM is primarily used for its representational capacity. The pipeline begins with \textit{model initialization and adaptation}, followed by \textit{data processing}. The PTM transforms input data into high-dimensional feature vectors or embeddings, which are consumed by downstream systems (\eg for clustering, retrieval, ranking). Either \textit{fine-tuning} or direct \textit{inference} may be applied to tailor the representations to the target task. Optional stages include \textit{feature engineering} and \textit{post-processing}.

\item \textbf{Generative-Oriented Pipeline.}
This organization centers on generating novel outputs such as text, code, or images. After \textit{model initialization, adaptation}, and \textit{data processing}, a defining stage is \textit{prompt}, which guides the model's generative behavior. Depending on the application, the PTM may be used as-is via \textit{inference}, or \textit{fine-tuned} for greater task alignment. \textit{Feature engineering} may be incorporated in hybrid methods (\eg retrieval-augmented generation). As in other organizations, optional stages such as \textit{post-processing} and \textit{delivery} serve to clean, format, or serve the outputs.

\item \textbf{Discriminative-Oriented Pipeline.}
This organization is structured for prediction tasks such as classification, regression, or structured prediction. After \textit{model initialization, adaptation}, and \textit{data processing}, the PTM is used as a predictive engine. Either \textit{inference} or \textit{fine-tuning} is used to make task-specific predictions. In some cases, lightweight \textit{feature engineering} enhances or augments model outputs. \textit{Post-processing} and \textit{delivery} stages handle formatting, thresholding, or integration into downstream services.
\end{enumerate}
\new{Across our dataset, each PTM is associated with one of these pipeline organizations. Feature extraction–oriented pipelines are the most common, accounting for 59\% of PTM pipelines, followed by discriminative-oriented pipelines at 21\%, and generative-oriented pipelines at 20\%.}

To complement the design-level characterization of reuse interactions, we also quantified the implementation-level organization of reuse pipelines across repositories. On average, a pipeline spans approximately 3.9 relevant source files per repository (median = 2), with each repository containing around 886 lines of relevant code in total (median = 361). Considering only repositories that reference more than one models, this corresponds to roughly 221 lines of code per model on average (median = 174).

\vspace{0.5em}

\FindingsBox{RQ2: \rqtwo}{Our analysis reveals that PTM reuse pipelines follow three dominant organizations: feature extraction, generative, and discriminative. Across all types, inference is the most consistently used stage, underscoring its central role in practical deployments. While some applications adopt minimal adaptation, complex workflows incorporate full-stage pipelines, including data processing, fine-tuning, post-processing, and delivery. The Model Adaptation stage is especially diverse; developers reuse PTMs as-is, initialize with pre-trained weights, add task-specific heads, or selectively modify layers. These patterns demonstrate that PTM reuse is rarely plug-and-play; instead, customization is key to optimizing performance and aligning with task-specific requirements across varied real-world scenarios.}

\subsection{\textbf{RQ3: \rqthree}}
\label{subsec:rq3}

\subsubsection{Methodology}
\label{subsec:rq3_methodology}

In RQ3, our goal is to understand how PTMs collaborate with other learned components, \ie scratch-trained models or additional PTMs, to form learned behaviors within ML-enabled systems.

Similar to RQ2, we first conduct a literature review to identify commonly discussed model interaction patterns across deep learning subfields, which forms a theoretical basis for our analysis. Next, we manually examine the 401 target projects and identify 200 that include more than one model; these form the dataset for RQ3. For each selected project, we analyze whether any PTM interacts with other model(s) within its reuse pipeline. We then map the observed interaction patterns to those identified in the literature and extend the taxonomy as novel patterns emerge. 

\paragraph{Step 1: Literature Review.} We begin by reviewing prior research to understand how model interactions are conceptualized in deep learning literature. Although our interest centers on PTMs, we consider both PTM and non-PTM interactions to develop a holistic view of reusable multi-model patterns.

\new{To collect relevant work, we focused on survey papers, as they summarize recurring design patterns and provide generalized taxonomies, rather than reporting isolated empirical findings. For this step, we relied exclusively on Google Scholar to identify highly cited studies on PTM workflow architectures and interaction patterns across diverse venues (e.g., software engineering, machine learning, and domain-specific conferences). Compared to venue-specific digital libraries, Google Scholar offers broader cross-disciplinary coverage and citation-based ranking, which aligns with our goal of capturing widely adopted reusable patterns. Although this strategy may miss some relevant work, we believe it is sufficient to cover commonly recognized patterns.} Similar to RQ2, we use the Publish or Perish tool to query Google Scholar with the following search string:

\begin{quote}
\hspace*{2em}\texttt{("deep learning" OR "large language model" OR "foundation \hspace*{2em} model" OR "pre-trained model" OR "pretrained model")}
\\
\texttt{AND "survey"}
\\
\texttt{AND ("interaction" OR "multi-modal" OR "multimodal" OR "multi-model" \hspace*{2em} OR "architecture" OR "fusion")}
\end{quote}

\new{We selected keywords based on deep learning terminology (deep learning, large language model, foundation model, pre-trained model) and interaction-related terms (interaction, multi-modal, multi-model, architecture, fusion). We focused on survey papers because they summarize recurring design patterns and provide generalized architectural taxonomies, rather than reporting isolated empirical findings. During scoping, we tested variations and found the final query balanced coverage and relevance well.}

This search yields the top 1,000 results ranked by citation count. From these, we select papers according to two criteria: (1) general-domain surveys (\eg NLP, computer vision, multimodal learning) that describe reusable architectures, and (2) domain-specific surveys (\eg robotics, medicine) from application areas where deep learning has seen significant adoption. 
Applying these criteria results in 224 survey papers.
Next, we apply adaptive reading~\citep{petersen2008systematic}, following the same approach used in RQ2, examining each survey paper in detail only when necessary to determine whether it explicitly discussed model interactions (\eg feature fusion, multi-PTM workflows). This process yields a final set of 145 survey papers.

\vspace{0.1cm}

After reviewing the selected papers, we find that model interactions are rarely defined or explicitly discussed in the literature, and that no prior work specifically addresses interactions between PTMs and other models (\S\ref{subsec:ptm-stages}). However, many studies implicitly describe recurring, high-level multi-model interaction patterns in their discussions of model design patterns or training strategies for specific domain applications. From these, we identify two forms of interaction between PTMs and other models:

\begin{itemize}
\item \textbf{Feature Handoff (FH).} This interaction occurs when the output of one model, typically a learned representation, is passed to another model for further processing. For instance, vision transformers often rely on convolutional backbones such as ResNet-50 to extract image features, which are then handed off to transformer layers for higher-level interpretation \citep{transformers_vision_survey}. Multimodal models frequently encode different modalities (\eg text and image) using separate specialized components, followed by fusion or decoding layers \citep{multimodal_llms_survey, fusion2024survey}. Similarly, in time series modeling, graph or recurrent encoders produce latent representations that are fed into task-specific heads \citep{liang2024foundation}. 

\item \textbf{Feedback Guidance (FB).} This interaction occurs when a model serves as an evaluator during the training stage or fine-tuning of another model. One model's outputs are assessed using similarity scores or perceptual distances computed by another model, and this signal is used to update the first model. For example, perceptual loss is often computed using pretrained vision networks such as VGG to shape image generation outputs \citep{oussidi2018deep}. In multimodal generative systems, pretrained embedding models (\eg CLIP) provide alignment feedback between text prompts and generated images, offering a form of self-supervised guidance without being updated themselves \citep{aigc2024survey, multimodal_image_editing_survey}. 
\end{itemize}

\paragraph{Step 2: Identify and Categorize Cross-stage Multi-model Interactions.} In this step, we examine how the model interactions identified in our literature review are manifested in real-world projects. We focus on detecting interactions between PTMs and other models, either additional PTMs or models trained from scratch, within the same project (\S\ref{subsec:ptm-stages}). 

From RQ2, we already have PTM reuse pipelines labeled with their functional stages (\eg feature engineering, inference, fine-tuning). For each labeled PTM reuse pipeline, we further examine whether additional models interact with it. For every identified interaction, we annotate the following: 

\begin{itemize}
    \item \textit{Stage}: The stage(s) at which the interaction occurs (\eg training, inference, evaluation) within the involved pipelines.
    \item \textit{Directionality}: Whether data flows into, out of, or through the PTM.
    \item \textit{Function}: The role of the PTM in the interaction, \eg providing features or embeddings to another model, or offering evaluative signals for the training of another model.
\end{itemize}

When interactions occur between two PTMs, the stages of both reuse pipelines are already labeled. However, when an interaction occurs between a PTM and a scratch-trained model, we must label the stage of the conventional ML pipeline (defined in Figure~\ref{fig:background_conv_pipeline}) in which the interaction takes place. We apply the same approach used for identifying PTM stages in real-world code, \ie leveraging ML-related frameworks to map the implementation details to the corresponding stages. 

For each interaction, we annotate its \textit{Stage}, \textit{Directionality}, and \textit{Function}. While stage and directionality describe the technical flow of data, the function captures the higher-level role of the PTM (\eg feature provider, evaluator, or refiner). We use function as the primary basis for categorization, resulting in four primary categories (Table~\ref{tab:ptm_interaction_types}): Feature Handoff, Feedback Guidance, Evaluation, and Post-Processing Refinement. The first author performs the initial labeling. Ambiguous or novel interaction types are discussed and resolved through consensus among all authors. This process yields a refined taxonomy grounded in real-world OSS practice.

\subsubsection{Result}

\noindent \textbf{Interaction Types in PTM Reuse Pipeline.}
Table~\ref{tab:ptm_interaction_types} summarizes the four interaction types identified in the 200 target projects. 
Feature Handoff and Feedback Guidance are the most common, each accounting for more than 40\% of observed examples, while Evaluation and Post-Processing are less frequent. This indicates that PTM reuse in OSS pipelines is dominated by patterns involving either direct feature passing or bidirectional feedback during training.
Compared with prior literature, which covers two interaction types without focusing on PTMs, our study identifies four interaction types, two more than previously recognized, and highlights finer-grained cross-stage PTM-specific patterns.

Next, we describe the characteristics and give examples for each category.

\begin{table}[t]
\centering
\caption{
Types of model interactions in PTM reuse pipelines.  ``Evaluation'' and ``Post-Processing Refinement'' represent additions to prior literature.
}
\renewcommand{\arraystretch}{1.3}
\resizebox{\textwidth}{!}{%
\begin{tabular}{@{}p{4cm} p{8cm} c}
\toprule
\textbf{Category} & \textbf{Description} & \textbf{Frequency} \\
\midrule
Feature Handoff & One model’s output is passed as structured input to another, forming multi-step pipelines. & 42\% \\
Feedback Guidance & A PTM provides evaluative signals (\eg CLIP loss) during training of another model. & 45\% \\
\textbf{Evaluation} & PTM used only for evaluation or metric computation (\eg FID, NER robustness), not during training or inference. & 9\% \\
\textbf{Post-Processing Refinement} & A PTM refines or validates the output of another model post-hoc. \eg safety checkers on generated images. & 4\% \\
\bottomrule
\end{tabular}
}
\label{tab:ptm_interaction_types}
\end{table}

\begin{figure}[h]
    \centering
    \includegraphics[width=0.9\linewidth]{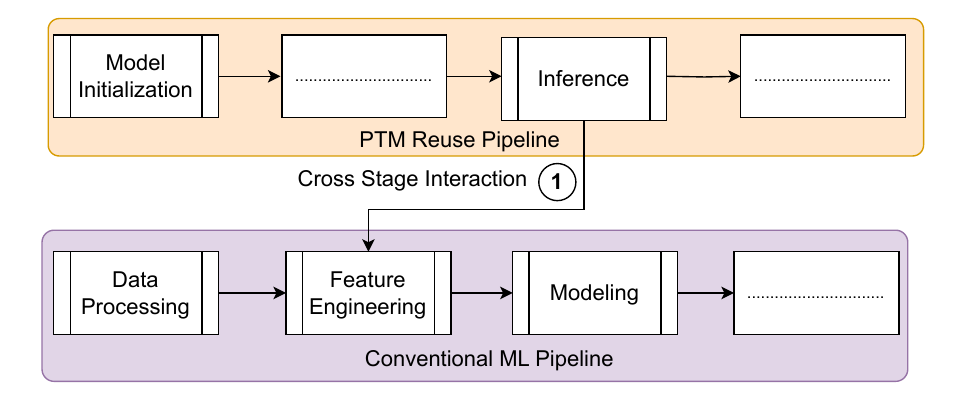}
    \caption{Feature Handoff: The PTM generates outputs during \textit{inference} and passes them (\ding{172}) to the \textit{feature engineering} stage of a downstream pipeline, enabling task-specific models to reuse pretrained representations.    
}
    \label{fig:heirarchical}
\end{figure}

\paragraph{Category 1: Feature Handoff.}
Feature Handoff describes scenarios where the output of a PTM is passed as structured or latent input to another model. A PTM can generate embeddings or intermediate representations through \textit{inference} or \textit{fine-tuning} and pass them to another pipeline's stage (\eg \textit{feature engineering}).\footnote{For example, see \url{https://github.com/zhvng/open-musiclm}} For example, a pretrained EfficientNet encoder can provide embeddings to a task-specific model, allowing it to leverage the PTM’s learned features without recomputing them. This hierarchical reuse of pretrained representations enables more efficient computation and integration across pipeline stages.
Figure~\ref{fig:heirarchical} illustrates this setup: output generated by the PTM in its reuse pipeline’s \textit{inference} stage is passed (Circle \ding{172}) to the \textit{feature engineering} stage of the conventional ML pipeline.

\begin{figure}[tp]
    \centering
    \includegraphics[width=0.9\linewidth]{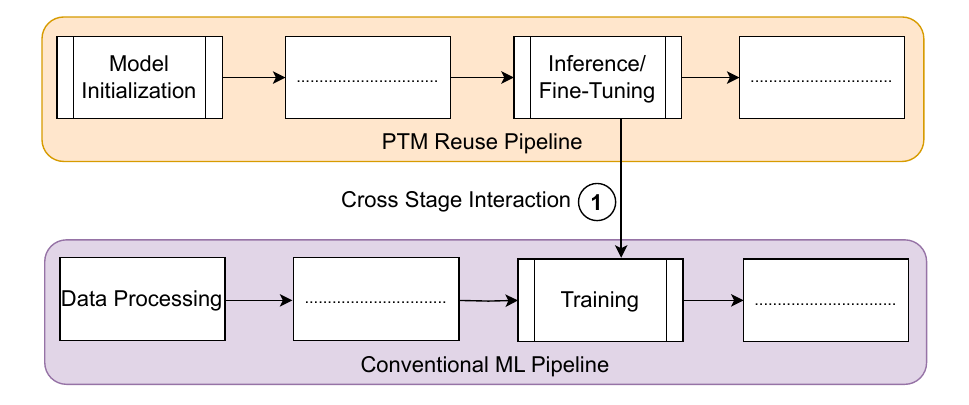}
    \caption{
    Feedback Guidance: The PTM performs \textit{inference} to produce outputs that are used to compute a loss, which is then passed \ding{172} to another pipeline and used during its \textit{training}.
    }
    \label{fig:feedback}
\end{figure}

\paragraph{Category 2: Feedback Guidance.}
In this interaction type, a PTM provides supervisory signals to another model during \textit{training} or \textit{fine-tuning}. These signals are often perceptual, semantic, or similarity-based, and are incorporated into custom loss functions to improve task-specific models. 
For example, a CLIP model computes similarity between text prompts and generated images. This similarity score is used as a perceptual loss to guide the image generator’s training.~\footnote{For example, see \url{https://github.com/ai-forever/Kandinsky-2}} 
The PTM may serve purely as a signal provider (Figure~\ref{fig:feedback}) or it may be trainable, with its parameters updated in tandem with those of the target model. In the latter case, a bidirectional feedback loop is formed.

\begin{figure}[tp]
    \centering
    \includegraphics[width=0.9\linewidth]{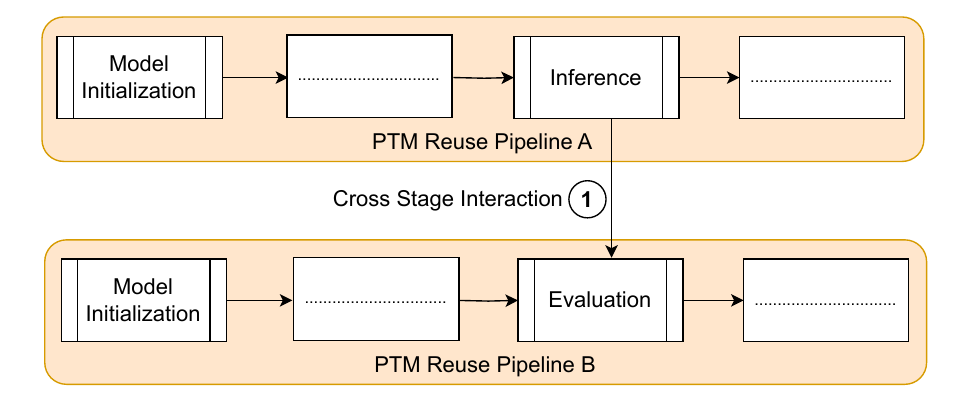}
    \caption{Post-Processing Refinement: The PTM A performs \textit{inference} to generate outputs that are used (\ding{172}) to evaluate PTM B's outputs during its \textit{evaluation} stage.}
    \label{fig:evaluation}
\end{figure}

\paragraph{Category 3: Evaluation.}
In this interaction type, a PTM's output is employed to assess the quality of another model's outputs. These interactions typically occur post hoc (at the evaluation stage). The PTM provides objective metrics or diagnostic insights but does not influence model behavior directly. This pattern is common in generative systems. For instance, Inception-v3 and CLIP are often used to compute Fréchet Inception Distance (FID) or CLIP-based similarity scores for evaluating images or text-to-image models. It also appears in NLP applications. For instance, pretrained language models (\eg GPT) can be used to generate paraphrased or alternative entity descriptions, which are then fed into a named entity recognition (NER) system to test its robustness. These models are not involved in the training or adaptation of the NER model but serve only as tools to generate diverse evaluation inputs.\footnote{For example, see \url{https://github.com/IBM/zshot}.}
Figure~\ref{fig:evaluation} illustrates a typical setup.

\paragraph{Category 4: Post-Processing Refinement.}

\begin{figure}[h]
    \centering
    \includegraphics[width=0.9\linewidth]{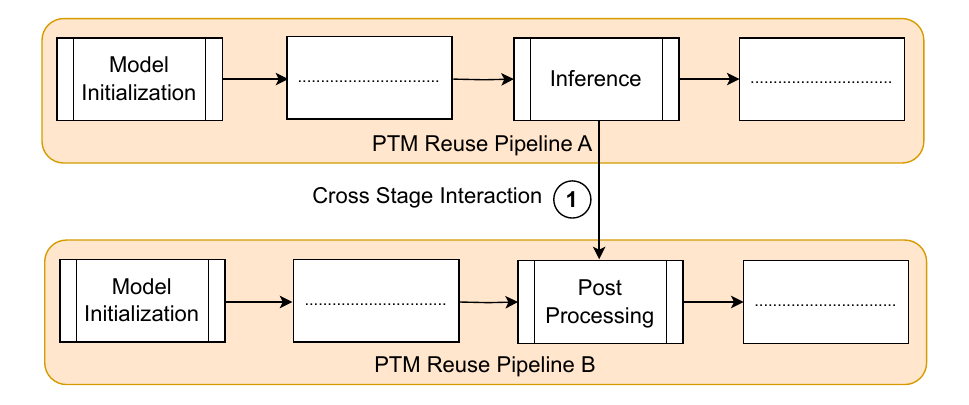}
    \caption{Post-Processing Refinement. The PTM A performs \textit{inference} to generate outputs that are used (\ding{172}) to refine PTM B's outputs during its \textit{postprocessing} stage.}
    \label{fig:postprocessing}
\end{figure}

In this interaction type, a PTM receives the \textit{original output} from another pipeline and performs inference to filter, rerank, or refine it during that pipeline’s \textit{post-processing stage}. The PTM does not contribute to generation or learning directly but serves as an auxiliary module that improves output quality or safety (Figure~\ref{fig:postprocessing}).
A common example is the use of a ``safety checker" PTM in Stable Diffusion pipelines, which filters generated images based on content moderation constraints.~\footnote{For example, see \url{https://github.com/IrisRainbowNeko/DreamArtist-stable-diffusion}} Similarly, in image captioning or text generation tasks, pretrained BERT models are sometimes used to rerank multiple candidate outputs based on fluency or semantic consistency.

\vspace{0.5em}
\noindent \textbf{Structural and Modular Characteristics of Interaction Types.} 
To gain deeper insights and compare the four interaction types, we qualitatively inspect a subset of examples from each category. Specifically, we review 20 Feature Handoff examples, 20 Feedback Guidance examples, 4 Evaluation examples, and 4 Post-Processing Refinement examples.
We analyzed all categories across three dimensions

\begin{itemize}
    \item \textbf{Graph Structure:} Topological arrangement of models;
    \item \textbf{Interaction Style:} The qualitative form of information exchange (\eg sequential handoff, feedback loop, or evaluative filtering). While related to stage-level directionality, which specifies whether data flows into, out of, or through a PTM, interaction style provides a higher-level description of the overall pattern.
    \item \textbf{Modularity:} Degree of coupling between models.
\end{itemize}

These dimensions are chosen because together they capture how models are arranged, how information flows, and how strongly components depend on each other. From this inspection, we observe several key findings.
\textit{Feature Handoff}, one of the most structurally diverse categories, can be further subdivided into three structures:
\begin{enumerate}[label=\textbf{1\alph*.}]
    \setlength{\itemindent}{2em} 
    \item \textbf{Hierarchical Processing} – PTMs stacked as feature extractors and downstream models (\eg encoder-decoder pipelines);
    \item \textbf{Sequential Processing} – Sequential pipelines where one model’s output feeds into the next (\eg T2I pipelines, QA systems);  
    \item \textbf{Parallel Collaborative Processing} – Multiple PTMs operate in parallel on different modalities or aspects, with outputs fused later (\eg audio-text fusion).
\end{enumerate}

These structural patterns differ in modularity: hierarchical processing (1a) shows tighter coupling due to stacked components, while sequential (1b) and parallel (1c) setups emphasize clean interfaces or modality-specific branches, promoting modularity.
\textit{Feedback Guidance} interactions (Type 2) involve backward-propagated supervisory signals during training, where the PTM may either provide only guidance or be trainable (co-updating with the target model), resulting in stronger coupling and lower modularity. In contrast, \textit{Evaluation} (Type 3) and \textit{Post-Processing} (Type 4) interactions are loosely coupled, operating independently of the core inference flow, making them highly modular and interchangeable. 
These findings highlight the functional and structural diversity of pretrained model reuse in OSS pipelines. Table~\ref{tab:interaction_structures} summarizes the structural, interaction style, and modularity characteristics of each type. Overall, these categories capture recurring architectural idioms used by OSS developers and reveal different PTM reuse modalities.

\vspace{0.2cm}

\begin{table}[ht]
\centering
\caption{
Structural, interaction style, and modularity characteristics of multi-model interactions in PTM reuse pipelines. Feature Handoff (Types 1a–1c) exhibits structural diversity, whereas other interaction types follow more consistent patterns.
}
\renewcommand{\arraystretch}{1.3}
\resizebox{\textwidth}{!}{%
\begin{tabular}{p{3cm}p{4cm}p{4cm}p{3cm}}
\toprule
\textbf{Interaction Type} & \textbf{Graph Structure} & \textbf{Interaction Style} & \textbf{Modularity} \\
\midrule
1a. Hierarchical Processing & Stacked modules forming deep pipelines (\eg encoder-decoder) & Tightly coupled feature passing across layers & Medium – stages are dependent and can be co-trained \\
1b. Sequential Processing & Directed acyclic flow across distinct pipeline stages & Stage-wise transformation; downstream consumes upstream output & High – clean interfaces enable reuse \\
1c. Parallel Collaborative Processing & Independent branches fused at a later stage & Parallel inference with late fusion & High – loosely coupled for multimodal or ensemble use \\
2. Feedback Guidance & Forward supervisory signals from PTM; optional backward flow if PTM is trainable & Supervisory signals propagate to downstream model during training; gradients may flow back if PTM is trainable & Medium–Low – strong coupling during co-training \\
3. Evaluation & No runtime coupling; outputs assessed post hoc & Passive metric reporting or evaluation by external module & High – completely decoupled \\
4. Post-Processing Refinement & Sequential append; refinement of existing outputs & Asymmetric reranking or validation without upstream access & High – plug-in style modules, loosely integrated \\
\bottomrule
\end{tabular}%
}
\label{tab:interaction_structures}
\end{table}

\FindingsBox{RQ3: \rqthree}{
We identified four types of interaction between PTMs and other learned components. The most prevalent pattern is \textit{Feature Handoff}, where PTMs pass intermediate outputs to another model in hierarchical, sequential, or parallel structures. Additionally, PTMs are used in \textit{Feedback Guidance} roles, providing non-trainable supervisory signals during training, as well as for \textit{Evaluation}, where PTMs act purely as metric calculators for another model. Finally, some PTMs serve in \textit{Post-Processing Refinement}, validating or modifying the outputs of other models. These interactions reveal a rich ecosystem in which multiple models collaborate under different levels of modularity, influencing the ways PTMs are reused.}

\section{Discussion and Implications}\label{sec:discussion}

This section reflects on our study’s findings from three perspectives. First, we discuss the broader conceptual implications for software dependencies, \new{focusing on the distinctions between Software Dependencies 1.0 and Software Dependencies 2.0}. Next, we examine the maintenance challenges that arise from PTM dependencies, pipeline complexity, and multi-model interactions in ML-enabled systems. Finally, we outline actionable implications for key ML stakeholders, including researchers, developers, PTM hub providers, and MLOps platform providers.

\subsection{Software Dependencies 1.0 vs. 2.0}
\label{subsec:comparison_dep}

\begin{table}[t]
\centering
\caption{Comparison of Software Dependencies 1.0 (code-centric) and Software Dependencies 2.0 (model-centric, PTMs).}
\label{tab:dependencies_2}
\renewcommand{\arraystretch}{1.5} 
\normalsize
\resizebox{\textwidth}{!}{%
\begin{tabular}{p{3.2cm} p{5cm} p{6cm}}
\toprule
\textbf{Aspect} & \textbf{Software Dependencies 1.0 (Code-Centric)} & \textbf{Software Dependencies 2.0 (Model-Centric, PTMs)} \\
\midrule

Dependency Structure & 
Static, uniform edges; "depends on," captured in manifests~\citep{wittern2016look}. & 
Semantic edges; PTM$\rightarrow$project (fine-tuning, adapters), PTM$\leftrightarrow$PTM (handoff, feedback). Meaning emerges from reuse pipelines and runtime context (\S\ref{subsec:rq2}, \S\ref{subsec:rq3}). \\

\new{Integration} &

\new{Often} simple; typically an \texttt{init()} function or small configuration file~\citep{decan2019empirical} &
\new{Often involves} high configuration cost; involves code API + model artifacts (\S\ref{subsec:rq1}), multiple parameters, adapter/head selection (\S\ref{subsec:rq2}), and multi-model coordination (\S\ref{subsec:rq3}). \\

Interchangeability \& Complementarity &
Rigid APIs~\citep{brito2018and}; substitution rare. & 
Substitution feasible within PTM families (\S\ref{subsec:rq1}); PTMs often combined across tasks/modalities, generating complex graphs (\S\ref{subsec:rq3}). \\

Behavior &
Predictable, deterministic (controlled randomness)~\citep{fowler2002patterns}. & 
Context-sensitive, emergent; outcomes shaped by reuse pipelines (\S\ref{subsec:rq2}), provenance, and cross-PTM interactions (\S\ref{subsec:rq3}). \\


Traceability &
Centralized, explicit manifests~\citep{decan2019empirical}. & 
Scattered, implicit; often hidden in configurations or runtime; no standard declaration (\S\ref{subsec:rq1}). \\

\bottomrule
\end{tabular}%
}

\end{table}


Our analysis highlights \new{notable differences} between conventional software dependencies (\textit{Software Dependencies 1.0}) and the paradigm of model-centric dependencies (\textit{Software Dependencies 2.0}). Our measurements echo another recent work by ~\citet{banyongrakkul2025release}, which emphasizes that integrating PTMs into downstream applications introduces unique challenges. We believe these challenges arise from a shift from code-centric to model-centric artifacts, motivating our introduction of the term Software Dependencies 2.0.
Table~\ref{tab:dependencies_2} summarizes these distinctions based on our findings. 
While Software Dependencies 1.0 are grounded in code-centric artifacts, the reuse of PTMs in ML-enabled systems introduces model-centric components whose behavior is shaped by the learned PTM, their reuse pipelines and interactions with other models. This evolution extends the unit of dependency management from a simple \textit{library + version} to a more complex composite: \textit{model + provenance + reuse pipeline + interaction}.


\subsubsection*{From Structural Edges to Semantic Edges}
In Software Dependencies 1.0, dependency graphs are static and uniform. Nodes represent libraries or packages, and edges denote a simple, purely structural ``depends on'' relation, typically declared in manifests such as \texttt{requirements.txt} or \texttt{pom.xml}~\citep{wittern2016look}. These edges are context-agnostic: they indicate that one unit cannot compile or run without another, but they do not alter program semantics. The graph’s meaning is fully determined by code and manifests.

By contrast, Software Dependencies 2.0 introduce semantic edges that carry meaning beyond simple structural linkage. For example, PTM$\rightarrow$project edges can take different forms depending on how the model is used: one edge type denotes that a PTM has been fine-tuned within the reuse pipeline, while another indicates that task-specific heads have been added (Table~\ref{tab:ptm_adaptation}). Similarly, PTM$\leftrightarrow$PTM edges can take different forms depending on the interaction pattern between two models: for example, one edge type denotes the handover of features, while another represents a feedback loop (Table~\ref{tab:ptm-usage-comparison}). These edges arise from reuse pipelines and runtime configurations, can shape system behavior, risks, and reproducibility rather than merely recording dependency existence.


\subsubsection*{\new{Integration}}

In Software Dependencies 1.0, code libraries often require minimal setup, typically a simple \texttt{init()} function or a small configuration file. In contrast, Software Dependencies 2.0 involve both code APIs and model artifacts (weights, configurations, adapters, etc.). PTM reuse requires managing multiple parameters, specifying model names with versions (Table~\ref{fig:breakdown}), selecting heads or adapters, adjusting configurations (Table~\ref{tab:ptm_adaptation}), and coordinating multi-model pipelines (Table~\ref{tab:interaction_structures}). This configuration complexity \new{suggests} that PTM dependencies are \new{typically} more complex to integrate than code-centric dependencies.


\subsubsection*{Interchangeability and Complementarity in PTM Reuse} In Software Dependencies 1.0, code-centric dependencies are largely independent. Substituting one dependency for another is rare~\citep{brito2018and}, and combining packages rarely produces emergent behavior. Interactions between dependencies are minimal.

In contrast, Software Dependencies 2.0 exhibit both interchangeability and complementarity in PTM reuse (Table~\ref{fig:ptm_dependency_tree}). Within the same family (\eg BERT variants), models can often be substituted with minimal code changes, reflecting a model-level Liskov substitution principle~\citep{liskov1994behavioral}. At the same time, PTMs are frequently combined across tasks or modalities, where one model's output serves as the other's input (Table~\ref{tab:interaction_structures}). These interactions generate complex graphs spanning code, data, and models, highlighting that 2.0 dependencies are qualitatively different from 1.0.

\subsubsection*{From Functional Interfaces to Behavioral Dynamics}

In Software Dependencies 1.0, system behavior is predictable and deterministic, fully determined by code semantics and versioned APIs~\citep{decan2019empirical,sammak2023developers}. Randomness is controlled and reproducible, for example via seeded random number generators.

By comparison, Software Dependencies 2.0 exhibit fluid, context-dependent behavior. Outcomes rely on reuse pipelines (Table~\ref{tab:merged_pipeline_stages}), fine-tuning, provenance, and interactions with other PTMs (Table~\ref{tab:ptm_interaction_types}). Emergent behaviors arise from the combinations of multiple PTMs, producing properties not present in individual models. Correctness, reproducibility, and safety of applications built on PTMs are therefore tied to how dependencies are composed within the ecosystem~\citep{gao2025ai}.


\subsubsection*{Engineering Gaps: Traceability and Management}
Traceability in Software Dependencies 1.0 is straightforward: manifests provide explicit, centralized dependency declarations~\citep{decan2019empirical}.
By comparison, in Software Dependencies 2.0, PTM usage is often implicit, hidden in configuration files, environment variables, or runtime calls (Figure~\ref{fig:breakdown}). Model hubs like Hugging Face serve as repositories but do not provide reproducible manifests or resolve a model’s internal dependencies. This opacity presents a pressing challenge for the discovery, governance, and integration of PTM dependencies.

\subsubsection*{\new{Debate: Does This Concept Merit the Label ``Software Dependencies 2.0''?}}
\label{subsec:debate}

\new{Our use of the term ``Software Dependencies 2.0'' is intended as an analytical abstraction rather than a categorical replacement of conventional dependency models. We do not claim that ML-enabled systems discard all properties of Software Dependencies 1.0, nor that every PTM integration exhibits all of the characteristics summarized in Table~\ref{tab:dependencies_2}. Instead, our conceptualization captures a recurring pattern observed across our empirical analysis: that model-centric artifacts introduce semantic dependency edges, pipeline-embedded adaptation, cross-model interaction, and decentralized traceability that are not fully captured by conventional manifest-based dependency models. We intend ``2.0'' to reflect a substantive extension of the unit of dependency management, from ``library + version'' to ``model + provenance + version + reuse pipeline + interaction''. Our extension is grounded in the structural regularities we observed across 401 projects.}

\new{We acknowledge that a reader might examine any of the rows in Table~\ref{tab:dependencies_2} and point to counterexamples in conventional software. For example, not all conventional dependencies are simple to incorporate, probabilistic algorithms are well known to introduce nondeterminism, and plugin ecosystems can exhibit complex inter-package interactions.
Our position follows the adage, ``\textit{The exception proves the rule}.''
While individual 1.0 systems may display isolated instances of these properties, our findings indicate that such characteristics are \textit{structural and pervasive} in PTM reuse.
We assert not that PTM reuse introduces entirely novel phenomena, but that it systematically combines behavioral, architectural, and provenance complexity in ways that make existing dependency abstractions incomplete.
We therefore offer the term Software~Dependencies~2.0 as a conceptual tool to reason about this emerging pattern, and its merit ultimately rests on whether it clarifies analysis and supports more reliable engineering practice.}

\subsection{PTM-Centric Maintenance Challenges in ML-Enabled Systems}
\label{subsec:debt}

While prior work has examined technical debt in ML systems~\citep{sculley2015hidden, bogner2021characterizing, gesi2022code, recupito2024technical}, our study \new{suggests} new maintenance challenges that arise from the shift from training models from scratch to reusing PTMs or combining PTMs with scratch-trained models. \new{Recent PTM-specific studies have cataloged practitioner challenges from different angles: \citet{tan2024challenges} focus on PTM usage issues such as model selection, testing, while \citet{banyongrakkul2025release} classify developer-reported challenges across seven themes, including model I/O handling, dependency incompatibility, hardware constraints, and missing model resources. These works characterize \textit{what} challenges developers encounter when working with PTMs. Our work complements theirs by examining \textit{how} these challenges manifest as maintenance concerns within reuse pipelines: specifically, through dependency structures and documentation practices (RQ1), pipeline-stage organization and adaptation patterns (RQ2), and cross-model interactions (RQ3).}


In Table~\ref{tab:techdebts}, we map the challenges categorized in our study to those reported in prior research on technical debt and other engineering challenges in ML-enabled systems. \new{We confirm maintenance-related aspects previously identified in conventional ML systems (\eg configuration complexity, dependency management) while extending them with PTM-specific and pipeline-centric insights from RQs. Notably, several maintenance challenges are specific to the PTM reuse context and have not been reported in prior work, including model-data entanglement and interaction debt from cross-stage multi-model coordination.} We then elaborate on each maintenance challenge category, linking it to our empirical findings.

\begin{table}[t]
\centering
\caption{
Conjectured technical debt in PTM reuse, illustrating maintenance challenges and example patterns observed in our study.
These challenges mirror existing software engineering concerns in conventional ML systems, such as configuration sprawl, but extend them in the context of PTM reuse.
}
\label{tab:techdebts}
\renewcommand{\arraystretch}{1.3}
\resizebox{\textwidth}{!}{%
\begin{tabular}{ p{3.65cm} p{3.0cm} p{4.0cm} }
\toprule
\textbf{\new{Maintenance} Challenge Category} & \textbf{Conventional ML (Prior Work)} & \textbf{PTM Reuse (This Study)} \\
\toprule
Dependency Management & Undeclared consumer dependencies~\citep{sculley2015hidden} & Informal PTM linking; missing version identifiers (\emph{dependency debt}) \\
 & Scattered parameters, configuration sprawl~\citep{bogner2021characterizing} & Uncontrolled distribution of PTM settings (\eg model name, version metadata) across files and scripts; absence of centralized tracking (\emph{configuration debt}) \\
 & Untracked seeds, environment drift~\citep{sculley2015hidden} & Missing metadata, model versions, or logs of PTM modifications (\emph{reproducibility debt}) \\
Adaptation & Data drift, model evolution~\citep{gesi2022code} & Custom model variants (\eg head swapping, adapters) without tooling or version tracking (\emph{adaptation debt}); silent model changes degrading downstream behavior (\emph{model drift debt}) \\
Pipeline Complexity & Glue code, pipeline jungle~\citep{sculley2015hidden, bogner2021characterizing} & Mixed generative/discriminative stages, diverse pipeline structures (\emph{pipeline debt}) \\
Model--Data Entanglement & --- & Outputs of one model serve as inputs to another; small changes in output format or content can break interacting pipelines (\emph{model–data entanglement}) \\
Multi-Model Coupling & Interface fragility, implicit coupling~\citep{bogner2021characterizing} & PTMs coordinate without clear handoff rules (\emph{interaction debt}) \\
Testing and Observability & Limited runtime observability, test brittleness~\citep{recupito2024technical} & Missing version-aware validation or debuggable intermediate outputs  \\
\bottomrule
\end{tabular}
}
\end{table}

\paragraph{Dependency Management.} Accurate and consistent PTM referencing is essential for effective dependency tracking, especially given that \ProjectsMultiPTM of projects in our dataset utilize multiple PTMs rather than a single one. In practice, however, PTM references are often scattered across source code, documentation, and configuration files, with only \ProjectsCentralizedDocumentation of projects declaring all PTMs outside of source code (\ie a README or 
configuration file)(\S\ref{subsec:rq1}). Additionally, only 12\% of projects specify the version of PTM. This decentralized and informal linking with missing version identifiers results in incomplete provenance metadata, creating \emph{dependency debt} that complicates the long-term management of model dependencies. PTM reuse also suffers from \emph{configuration debt}, \ie the uncontrolled scattering of model metadata across multiple scripts, configuration files, and code blocks (\S\ref{subsec:rq1}), and from \emph{reproducibility debt}, as downstream behavior can change unexpectedly when upstream models are updated due to the absence of structured metadata such as model names and versions. These findings align with prior work on undeclared consumer dependencies in ML systems~\citep{sculley2015hidden} 
\new{However, we acknowledge that PTM versioning is not commonly reported as a critical 
practitioner challenge in existing taxonomies~\citep{banyongrakkul2025release,tan2024challenges}, suggesting it may represent a latent maintenance risk rather 
than an immediate usability concern for many developers.}

\paragraph{Adaptation.}  
PTM reuse also involves diverse adaptation strategies, such as adding task-specific heads, swapping or pruning layers, and integrating adapters (\S\ref{subsec:rq2}). These modifications are frequently performed without formal tooling, documentation, or version tracking, leading to \emph{adaptation debt} that hinders rollback and validation. Silent updates to pretrained models, such as architectural changes or retraining with different data, can further degrade downstream performance over time, resulting in \emph{model drift debt}. Together, these issues underscore that PTM reuse is rarely plug-and-play and requires systematic change management practices.

\paragraph{Pipeline Complexity.}  
PTM reuse pipelines often span multiple stages and can involve several distinct pipelines that interact or overlap, frequently chaining PTMs across feature extraction, generative, and discriminative tasks. Such cross-stage coordination amplifies architectural complexity, producing \emph{pipeline debt}. These pipelines exhibit diverse graph structures, including hierarchical stacks, sequential flows, and parallel branches (Table~\ref{tab:interaction_structures}). Hierarchical pipelines create deep, interdependent module chains, sequential pipelines enforce strict stage-wise dependencies, and parallel branches require careful fusion or synchronization of multiple outputs. These structural patterns complicate testing and maintenance, mirroring prior reports of glue code and “pipeline jungles”\citep{sculley2015hidden,bogner2021characterizing}, but are further amplified in systems that use PTMs, where models serve simultaneously as data processors and code modules.


\paragraph{Model–Data Entanglement.} Multiple forms of interaction between PTMs and other models within ML pipelines—such as feature handoff, feedback, evaluation, and post-processing refinement (\S\ref{subsec:rq3}). Downstream data depends directly on the outputs of upstream models. These dependencies create \emph{model–data entanglement}, where minor changes in upstream output structure or content can silently break downstream components.


\paragraph{Multi-Model Coupling.} PTMs participate in diverse multi-model coordination patterns with varying graph structures, signal flows, and degrees of modularity (\S\ref{subsec:rq3}). Hierarchical feature handoffs exhibit strong coupling; sequential and parallel handoffs are more modular; feedback loops introduce bidirectional dependencies that can erode modularity; and evaluation or post-processing stages remain loosely coupled but often interchangeable. However, these multi-model assemblies are implicit and lack representation schemas, leading to \emph{interaction debt} that reduces composability and hinders safe evolution.

\paragraph{Testing and Observability.} 
While we did not systematically study testing, our analysis suggests that PTM reuse across multiple pipelines(\S\ref{subsec:rq2}) and cross-stage interactions(\S\ref{subsec:rq3}) exacerbate monitoring and debugging difficulties. Such complexity makes root cause analysis challenging and contributes to a form of technical debt in observability, where multi-pipeline coordination increases brittleness and reduces transparency.


The above maintenance challenges suggest a need to revisit traditional software engineering abstractions, such as modules, interfaces, and configuration boundaries, in the context of PTM-integrated ML systems. PTM reuse pipelines exhibit dynamic integrations that blur boundaries between data, code, and models. These factors challenge established principles of modular design, testing, and interface stability, especially as models evolve outside direct developer control. The rise of multi-model, PTM-based architectures introduces novel sources of complexity, coupling, and debt.

\subsection{Implications}

Our study of PTM reuse pipelines reveals growing complexity in the engineering of ML-enabled systems that incorporate PTMs. Based on findings from RQ1–RQ3, we outline implications for four key groups: researchers, developers, PTM hub providers, and MLOps platform providers.

\paragraph{Researchers.} 
Our study contributes new knowledge and taxonomies derived from prior literature and qualitative analysis of real-world software projects that reuse PTMs. These findings reveal multiple opportunities for advancing methods and architectures to better support ML-enabled systems in practice.

PTM reuse introduces multiple forms of technical debt (\S\ref{subsec:debt}), including dependency, configuration, adaptation, and pipeline debt. These debts highlight gaps where future SE research can make a tangible impact. One priority is developing robust dependency management approaches tailored for PTMs, such as integrating ML-specific metadata into software bill-of-materials (SBOM) standards~\citep{bennet2025implementing}, enabling traceable records of model versions, fine-tuning settings, and lineage. Standardized model lock files and reproducible build procedures are also critical for supporting reproducibility, vulnerability tracing, and risk assessment~\citep{stalnaker2025ml}. For Software Dependencies 2.0, vulnerabilities propagate differently from code flaws~\citep{nist2024_ai_risk}, and current vulnerability databases for models are limited or outdated~\citep{mitre_attack,avidml2023}. 
Our findings indicate that vulnerabilities or attacks in PTMs can transfer downstream through reuse pipelines and multi-model interactions, highlighting the need for systematic methods to detect, evaluate, and mitigate these risks across cross-stage model pipelines.

Even in traditional systems, dependencies can be nuanced, distinguishing between included, forked, or patched code\citep{cisa2025_sbom}. PTMs amplify this complexity: fine-tuned or adapted models introduce derived relationships that are not explicitly tracked, creating challenges for provenance, versioning, and reproducibility. This complexity is further illustrated by tools and frameworks~\citep{jiang2023naming,duan2024modelgo}, which attempt to capture lineage and relationships among models. Together, these examples show that while conventional dependencies already involve subtle interconnections, PTM reuse significantly magnifies these challenges, underscoring the need for systematic metadata, tooling, and process support to manage Software Dependencies 2.0.

Another important direction is extending software quality frameworks~\citep{gesi2022code,studer2021towards} to account for the unique PTM lifecycle. Attributes such as fairness, interpretability, and robustness can evolve as PTMs are reused in pipelines~\citep{bender2021dangers}. Systematic measurement and management of these qualities will enable new evaluation methodologies and engineering practices.
Additionally, Software Dependencies 2.0 highlights future research opportunities in managing interchangeable and complementary PTM dependencies, including safe model substitution, coordination across heterogeneous models, and designing frameworks that support both substitution and composition in complex pipelines.

Finally, our results suggest complementary opportunities for ML
research, including methods for fine-tuning with clear
provenance~\citep{meiklejohn2025machine}, model watermarking to track
PTM lineage across
reuse~\citep{xian2024raw,cohen2025watermarking},
model-editing strategies, and metrics for multi-model pipeline
evaluation.

Together, these directions provide an empirically grounded agenda for both SE and ML research communities, promoting more sustainable, reproducible, and trustworthy reuse of pretrained models in software systems. We provide a foundational base, while future work should investigate data–model–code entanglement across cross-stage model interactions.

Future work should also examine PTMs accessed via model-centric platforms (MCPs)~\citep{krishnan2025advancing}, agent hubs~\citep{ehtesham2025survey}, and commercial APIs~\citep{openai2023gpt4}.
\new{While PTM reuse provides a canonical instantiation of Software Dependencies 2.0, other model-centric artifacts, such as hosted foundation models, retrieval-augmented pipelines, adapter libraries, and agentic compositions, may exhibit similar structural properties.
These environments may surface different dependency declaration practices, provenance mechanisms, and interaction patterns compared to open-source hubs.
Analyzing such systems would help assess the generality of Software Dependencies 2.0 as an abstraction and refine its boundaries across broader ML software ecosystems.}

\paragraph{Developers.} Our analysis highlights maintenance challenges that arise when PTMs are integrated into projects. Developers should adopt structured practices to manage this integration. Models should be consistently declared, version-pinned, and treated as evolving software components to support dependency tracking and versioning~(\S\ref{subsec:rq1}). Adaptations such as task-specific head additions or layer modifications should be documented across pipeline stages, and automated tools could help generate such documentation (\eg visualizations of model changes, integration points, and interfaces with other components) based on PTM reuse patterns and source code~(\S\ref{subsec:rq2}). Integration tests should be included in CI/CD pipelines to ensure reliability and consistency. Outputs from multi-model interactions should be monitored to prevent downstream failures~(\S\ref{subsec:rq3}). Standardized tooling for dependency management, pipeline testing, and performance monitoring can further support reproducibility and stability in PTM-based systems.

\paragraph{PTM Hub Providers.}
Our study highlights gaps in managing Software Dependencies 2.0. Traditional package managers and dependency graphs are insufficient for PTM reuse, and existing PTM ``managers'', such as Hugging Face, offer only partial support: they lack standardized manifest files, reproducible builds, and complete visibility into model versions or third-party library dependencies, leaving downstream projects with incomplete dependency information.
To help dependency management, hubs can provide immutable, versioned snapshots, structured changelogs, and rich metadata capturing model provenance. 
RQ2 (\S\ref{subsec:rq2}) showed diverse adaptation stages across three pipeline types, emphasizing that PTM reuse is rarely plug-and-play. Hubs can support developers by offering tools to visualize or compare model differences, highlight adaptation changes, and track compatibility across versions, and allow downloads of model components. Integration with CI/CD can help with automated testing of adapted models, dependency checks, and safe deployment of PTM updates.
Finally, RQ3 (\S\ref{subsec:rq3}) demonstrated frequent cross-stage interactions. Hubs can provide metadata or APIs exposing interaction patterns, helping developers anticipate and manage potential incompatibilities in complex, multi-model pipelines.
Together, these observations highlight that dedicated tooling for PTM dependency management should be developed.

\paragraph{MLOps Platform Providers.}
MLOps systems must become PTM-aware, treating the pipeline itself as a first-class component, extracting, tracking, and validating model metadata during execution—including version, adaptation and fine-tuning history. To manage pipeline complexity (\S\ref{subsec:rq2}), platforms can provide metrics for interaction quality, such as reuse intensity, adaptation overlap, and dependency distance. Observability mechanisms help track and validate PTMs within complex pipelines, making it easier to detect issues when models interact or when downstream data depends on upstream outputs~\citep{pineau2021improving}. Integrating these checks into CI/CD workflows will allow automatic downstream validation whenever a PTM is swapped or updated, supporting more robust and maintainable multi-PTM pipelines and cross-stage interactions (\S\ref{subsec:rq3}). Such automated workflow enforcement resonates with challenges reported in Workflows as Code systems, where defining, executing, and monitoring complex pipelines also require systematic observability and validation practices~\citep{yasmin2024empirical}.

\section{Threats to Validity}\label{sec:threats}

\subsection{Construct Validity}
\label{subsec:construct_validity}

Construct validity concerns whether the measurements and procedures used in the study accurately reflect the theoretical concepts being investigated.

In our study, we examine patterns of PTM reuse in open-source software. To represent this concept, we used an existing and previously validated dataset, which includes OSS projects known to incorporate PTMs from Hugging Face and PyTorch Hub. The use of a curated dataset improves construct validity by relying on definitions and inclusion criteria already accepted in the research community.

However, we acknowledge that any dataset, regardless of its acceptance, embeds certain assumptions. For example, it may focus on specific types of PTMs, code structures, or documentation practices, potentially overlooking more implicit or unconventional forms of reuse. While our use of this dataset supports consistency and reproducibility, it may still omit edge cases or emerging practices not captured at the time of its creation.

\subsection{Internal Validity}
\label{subsec:internal_validity}

Internal validity refers to the degree to which the observed results can be attributed to the methods used rather than to other factors.

Our study involved a manual analysis of OSS projects from Hugging Face and PyTorch Hubs to identify patterns in PTM usage. To mitigate the risk of sampling bias, we used a systematic sampling strategy from the PeaTMOSS dataset (\S\ref{subsec:data_collection}), ensuring coverage across multiple domains, languages, and project sizes. Manual annotation introduces the possibility of bias, but we addressed this by employing inter-rater agreement to ensure consistency and reliability in the annotations.

Although \texttt{git grep} is a fast and effective way to locate references, it can produce false positives (\eg matches in comments or unrelated variables) and false negatives (\eg due to aliasing, indirection, or dynamic constructs) (\S\ref{subsec:data_collection}). To mitigate these issues, we manually verify matches for correctness and extend our inspection to common usage patterns of PTM loading and related API calls. While this improves coverage, some dynamic or heavily abstracted usages may still be missed, which we acknowledge as a limitation.

\subsection{External Validity}
\label{subsec:external_validity}

External validity concerns the extent to which our findings generalize to other contexts beyond the studied sample.

A primary consideration is our reliance on the PeaTMOSS dataset, which includes OSS projects created up until July 2023, with our analysis based on the February 2024 snapshot. The dataset predominantly uses PTMs from Hugging Face and PyTorch Hub. 
While these hubs represent the largest open repositories of PTMs at the time, they do not fully capture the reuse of other important families of models distributed via commercial APIs (e.g., OpenAI, Anthropic) and specialized model registries (e.g., NVIDIA NGC). In particular, our data set underrepresents PTMs that are distributed through closed, commercial, or restricted access registries, which may follow different reuse practices than the open source hubs captured in PeaTMOSS.

We chose PeaTMOSS because it provides a curated, large-scale, and publicly accessible dataset, enabling reproducible analysis. Since PTM technology and usage patterns evolve rapidly—with advances in architecture, training methods, and dependency practices—our findings may become outdated.

However, because our study focuses on general patterns of PTM reuse, it is not domain-specific and captures design principles applicable across diverse software projects, making the results broadly generalizable. Future studies should revisit these trends over time and extend the analysis to newer projects, additional domains, and different stages of software development to assess stability and generality.

\new{\subsection{Conclusion Validity}}
\label{subsec:conclusion_validity}
\new{Conclusion validity concerns whether our observed relationships and interpretations are well supported by the data. For project categorizations and the investigation of RQs, our taxonomies were developed through systematic sampling and manual analysis, which may not capture the full diversity of PTM usage patterns and inherently involve subjective judgment. In particular, for RQ2 and RQ3, our systematic review process required manual inspection and categorization of code patterns to identify stages and interaction types, introducing potential researcher bias.}

\new{To mitigate these threats, we adopted several safeguards. First, we developed explicit coding guidelines with representative examples for each category. Second, we conducted inter-rater reliability checks, with multiple researchers independently coding subsets of the data. Third, we held regular meetings to discuss disagreements and resolve ambiguities. Fourth, we iteratively refined our taxonomies as new patterns emerged during analysis. When applying the taxonomies to larger samples, we further validated them through pilot applications and documented edge-case resolutions (\S~\ref{subsec:data_collection}, \S~\ref{subsec:rq1}, \S~\ref{subsec:rq2}, \S~\ref{subsec:rq3}). Despite these safeguards, subjective judgment remains inherent in qualitative analysis, and alternative interpretations of certain code patterns are possible.}

\new{Finally, our interaction taxonomy is grounded in patterns observed within our dataset and may not be exhaustive. For example, while we identified parallel feature handoff patterns, we did not observe ensemble-style mechanisms (\eg voting or averaging) in our sample, which may reflect our dataset's characteristics rather than the absence of such patterns in practice.}

\new{Our versioning analysis examined multiple sources beyond just model-loading statements, including configuration files (YAML/JSON), README files, and requirements specifications. However, we acknowledge that our analysis did not extend to all possible version specification channels such as Docker images, Conda environments, or CI/CD configurations.}

\new{Our categorization of multi-PTM relationships uses high-level categories (interchangeable, complementary); finer-grained subcategories based on architecture, modality, or functional role could reveal additional patterns, but were not pursued due to concerns about coding complexity and inter-rater reliability. Future work could explore finer-grained subcategories of complementary PTM relationships based on architecture, modality, and functional role to identify potential connections between these dimensions.}

\new{Our comparison between Software Dependencies 1.0 and 2.0 is based on empirical observations from open-source projects rather than direct practitioner validation. Some challenges, such as versioning, may represent latent risks rather than immediate practical concerns. A user study validating these findings from a practitioner perspective is an important direction for future work.}

\vspace{0.5em}

\section{Related Work}\label{sec:related_work}
The widespread adoption of PTMs has created a complex software ecosystem, which researchers are beginning to analyze as a PTM supply chain~\citep{jiang2022empirical,stalnaker2025ml}, starting from upstream model hubs and packages and extending to their adaptation and integration within downstream software projects. While our work focuses on the downstream side of this chain, we review both perspectives to situate our contribution. We acknowledged some prior work in \S\ref{sec:background}, \S\ref{sec:discussion} and revisit relevant lines of research here, focusing first on hub-oriented (``upstream'') perspectives before turning to open-source ML-enabled systems (``downstream'') studies.

\subsection{Model Hubs and Upstream Perspectives}
A growing body of research examines the distribution, dependencies, and management of PTMs via model hubs, with a focus on the upstream aspects~\citep{jiang2022empirical,taraghi2024deep,castano2023model,taraghi2024deep,ajibode2025towards,jones2024we}. Jiang \etal~\citep{jiang2022empirical, jiang2024peatmoss} conducted pioneering work by empirically examining security vulnerabilities in popular model hubs and curating large-scale datasets to enable further research on PTM dependencies. Additional studies have investigated Hugging Face as a central hub, exploring evolution~\citep{castano2023model}, challenges~\citep{taraghi2024deep}, naming practices~\citep{jiang2023naming}, semantic versioning~\citep{ajibode2025towards}, licensing landscapes~\citep{stalnaker2025ml}, and systematically comparing qualitative and quantitative characteristics of PTMs~\citep{jones2024we}.
Beyond hub-focused research, work has also examined upstream practices at the package level, analyzing how PTLMs are trained, packaged, and released, providing insights into the preparation and dissemination of models before they reach hub platforms~\citep{adekunle2025synchronization}.

These works primarily focus on upstream packages and hubs and do not investigate downstream source code integration, adaptation, or practical usage within real software projects. Our work addresses this gap by moving beyond hub-centric analysis to empirically characterize downstream PTM reuse patterns, examining how models are integrated, adapted, and interacted with in actual software systems.

\subsection{Model Integration in ML-Enabled Systems}
Recent research has analyzed integration and usage trends of ML models in open-source systems, often considering all ML models rather than focusing on PTMs~\citep{sens2024large,11029840}.
For instance,~\citet{sens2024large} analyzed 2,928 ML-enabled systems on GitHub, finding that ML is primarily used for technical tasks rather than end-user functionality.
Over half of the projects were ML libraries, with only 18\% targeting end users.
Extensive model reuse was observed, often via code duplication across repositories, introducing dependency challenges.
However, the manual qualitative analysis was limited to 26 sampled systems, restricting the generalizability of insights into model reuse and model interactions.
Importantly, the dataset was not curated specifically for PTM reuse.
Meanwhile,~\citet{11029840} study curated 262 open-source ML products and analyzed 30 representative cases, revealing distinctive architectural patterns, limited involvement of data scientists, and minimal adherence to best practices such as testing, monitoring, and pipeline automation. While these findings provide insights into production-oriented ML products, they do not focus on PTMs or cross-stage model reuse. Our distinct focus allowed us to examine PTM-specific reuse, addressing gaps in prior work.

\new{
Prior work has examined the broader landscape of ML library usage and software engineering challenges in ML-enabled systems.
\citet{dilhara2021understanding} studied ML library usage and evolution, analyzing how developers interact with ML libraries and how these libraries evolve over time. In the context of what~\citet{karpathy2017software2} termed ``Software 2.0,'' their work provides insights into library-level dependencies and API usage patterns. While their focus is on ML libraries broadly, our work specifically targets PTMs as a distinct form of dependency, examining not just library usage but the full integration pipeline, including model-specific stages (\eg fine-tuning, inference) and cross-stage interactions between PTMs and other components.
More recently,~\citet{shivashankar2025maintainability} surveyed maintainability and scalability challenges in ML systems, identifying pain points and solutions. While their work provides a comprehensive overview of ML system challenges, our study contributes empirical evidence specifically for PTM dependencies—characterizing how these dependencies are structured (RQ1), organized within reuse pipelines (RQ2), and composed with other models (RQ3). Our PTM-specific focus reveals patterns and maintenance challenges unique to model-centric dependencies that may not fully emerge in broader ML system studies.}

A growing line of research has examined the reuse of PTM specifically in downstream software projects~\citep{jiang2024peatmoss, duan2024modelgo, banyongrakkul2025release}.
For instance,~\citet{banyongrakkul2025release} examined PTM reuse specifically in downstream software projects by analyzing GitHub issues and pull requests.
Their taxonomies recorded challenges in model usage, adaptation, configuration, and resource management, showing that PTM-related issues take longer to resolve than non-PTM issues. While these studies highlight barriers, they do not capture systematic PTM integration within software systems. Other research has documented metadata for 281,638 PTM packages and 28,575 downstream repositories, including license information, revealing that 0.24\% of licenses between PTMs and their dependent projects are inconsistent~\citep{jiang2024peatmoss}. Tools like ModelGo complement this by assessing license conflicts, improper license choices, and obligations across software, datasets, and PTMs~\citep{duan2024modelgo}. These findings highlight the gaps in understanding downstream adoption and reuse of PTM that we work to address in this work.

Unlike prior small-scale mixed-model datasets—covering both traditional ML models and PTMs~\citep{sens2024large,11029840}, PeaTMOSS~\citep{jiang2024peatmoss} enables detailed analysis of PTM dependencies, supporting structured insights into PTM integration. Based on this dataset, our work aims at PTM-specific reuse, combining large-scale mining with qualitative analysis to systematically categorize reuse types (RQ1), map pipeline stage utilization (RQ2), and investigate cross-stage model interactions (RQ3). This approach fills a critical gap in the literature by focusing exclusively on PTM reuse, rather than general ML model usage, offering actionable insights for improving reproducibility, maintainability, and design of ML-enabled systems.

Recent work on the LLM supply chain highlights the downstream application ecosystem as a critical stage where pre-trained models are operationalized in real-world applications~\citep{wang2025large,hu2025large}. While these contributions provide conceptual insights, they remain largely visionary and non-empirical. In contrast, our work focuses on systematic, empirical analysis of PTM reuse and integration in downstream software projects, addressing real-world adoption patterns, pipeline stage usage, and cross-stage model dependencies, thereby complementing and grounding the conceptual findings in prior LLM supply chain research.

\vspace{0.5em}

\section{Conclusion}\label{sec:conclusion}
This study provides the first large-scale empirical characterization of pre-trained models (PTMs) as a new form of software dependency, which we term \textbf{\textit{Software Dependencies 2.0}}. By analyzing a representative sample of 401 GitHub projects, we examined how PTMs are managed and integrated into real-world software. Our results show that PTM dependencies are often inconsistently specified across code, documentation, and configuration files, and adaptation is common, ranging from as-is reuse to architectural modification. We identified three dominant PTM reuse pipeline types --- feature extraction, generative, and discriminative --- and observed frequent interactions among PTMs and with other models, introducing significant architectural complexity beyond conventional library reuse. \new{Taken together, these findings provide empirical grounding for Software Dependencies 2.0 as model-centric, pipeline-embedded, and interaction-driven dependencies.}

\new{
Our results suggest that managing Software Dependencies 2.0 requires practices that extend beyond traditional dependency management, including improved traceability of model artifacts, explicit versioning, and support for multi-model coordination.
By conceptualizing PTMs as first-class, modular components rather than opaque add-ons, 
the software engineering community can develop tools, standards, and architectural guidance tailored to model-centric systems.
More broadly, advancing the theory and engineering of Software Dependencies 2.0 will be critical to support reliable integration and long-term maintenance as model-centric components become increasingly central to modern software systems.
Our results guide future research: 
  to investigate automated support for PTM dependency management,
  examine reuse and adaptation in industrial-scale ML systems,
  and
  further explore multi-model interactions to advance understanding and engineering of Software Dependencies 2.0.
}


\clearpage
\begin{sloppypar}
\bibliographystyle{elsarticle-harv}
\bibliography{main}
\end{sloppypar}

\end{document}